\RequirePackage[final]{graphicx}  
\documentclass[%
 reprint,
showpacs,preprintnumbers,
 amsmath,amssymb,
 aps,
 superscriptaddress,
prd,
floatfix,
draft
]{revtex4-1}

\usepackage{siunitx}
\usepackage{bm}

\usepackage[caption=false]{subfig}

\usepackage{units}
\usepackage{xspace}
\usepackage{placeins}
\usepackage{hyphenat}

\def\geant/{\textsc{geant4}}
\def\genie/{\textsc{genie}}
\def\ppfx/{\textsc{ppfx}}

\newcommand{\nova}{NOvA\xspace}
\newcommand{\numu}{\ensuremath{{\nu}_{\mu}}\xspace}
\newcommand{\numubar}{\ensuremath{\bar{\nu}_{\mu}}\xspace}
\newcommand{\nue}{\ensuremath{{\nu}_{e}}\xspace}
\newcommand{\nuebar}{\ensuremath{\bar{\nu}_{e}}\xspace}
\newcommand{\nutau}{\ensuremath{{\nu}_{\tau}}\xspace}
\newcommand{\nuetonumu}{\ensuremath{\nu_e \rightarrow \nu_\mu}\xspace}
\newcommand{\numutonue}{\ensuremath{\nu_\mu \rightarrow \nu_e}\xspace}
\newcommand{\numutonumu}{\ensuremath{\nu_\mu \rightarrow \nu_\mu}\xspace}

\newcommand{\nueCC}{\ensuremath{\nu_e\,{\rm CC}}\xspace}
\newcommand{\numuCC}{\ensuremath{\nu_{\mu}\,{\rm CC}}\xspace}
\newcommand{\nutauCC}{\ensuremath{\nu_{\tau}\,{\rm CC}}\xspace}

\newcommand{\dmsq}[1]{\ensuremath{\Delta m^2_{ #1 }}\xspace}
\newcommand{\sinsqtwo}[1]{\ensuremath{\sin^{2}2\theta_{ #1 }}\xspace}
\newcommand{\sinsq}[1]{\ensuremath{\sin^{2}\theta_{ #1 }}\xspace}
\newcommand{\evsq}{\ensuremath{{{\rm eV}^2}/c^4}\xspace}
\newcommand{\dcp}{\ensuremath{\delta_{\rm CP}}\xspace}

\newcommand{\anue}{\mbox{$\overline{\nu}_{e}$}\xspace}                   
\newcommand{\anumu}{\mbox{$\overline{\nu}_{\mu}$}\xspace}                

\begin{document}

\preprint{FERMILAB-PUB-18-223-ND}

\title{New constraints on oscillation parameters from \nue appearance and \numu disappearance in the \nova experiment}

\newcommand{\ANL}{Argonne National Laboratory, Argonne, Illinois 60439, 
USA}
\newcommand{\ICS}{Institute of Computer Science, The Czech 
Academy of Sciences, 
182 07 Prague, Czech Republic}
\newcommand{\IOP}{Institute of Physics, The Czech 
Academy of Sciences, 
182 21 Prague, Czech Republic}
\newcommand{\Atlantico}{Universidad del Atlantico,
Km. 7 antigua via a Puerto Colombia, Barranquilla, Colombia}
\newcommand{\BHU}{Department of Physics, Institute of Science, Banaras 
Hindu University, Varanasi, 221 005, India}
\newcommand{\UCLA}{Physics and Astronomy Department, UCLA, Box 951547, Los 
Angeles, California 90095-1547, USA}
\newcommand{\Caltech}{California Institute of 
Technology, Pasadena, California 91125, USA}
\newcommand{\Cochin}{Department of Physics, Cochin University
of Science and Technology, Kochi 682 022, India}
\newcommand{\Charles}
{Charles University, Faculty of Mathematics and Physics,
 Institute of Particle and Nuclear Physics, Prague, Czech Republic}
\newcommand{\Cincinnati}{Department of Physics, University of Cincinnati, 
Cincinnati, Ohio 45221, USA}
\newcommand{\CSU}{Department of Physics, Colorado 
State University, Fort Collins, CO 80523-1875, USA}
\newcommand{\CTU}{Czech Technical University in Prague,
Brehova 7, 115 19 Prague 1, Czech Republic}
\newcommand{\Dallas}{Physics Department, University of Texas at Dallas,
800 W. Campbell Rd. Richardson, Texas 75083-0688, USA}
\newcommand{\Delhi}{Department of Physics and Astrophysics, University of 
Delhi, Delhi 110007, India}
\newcommand{\JINR}{Joint Institute for Nuclear Research,  
Dubna, Moscow region 141980, Russia}
\newcommand{\FNAL}{Fermi National Accelerator Laboratory, Batavia, 
Illinois 60510, USA}
\newcommand{\UFG}{Instituto de F\'{i}sica, Universidade Federal de 
Goi\'{a}s, Goi\^{a}nia, Goi\'{a}s, 74690-900, Brazil}
\newcommand{\Guwahati}{Department of Physics, IIT Guwahati, Guwahati, 781 
039, India}
\newcommand{\Harvard}{Department of Physics, Harvard University, 
Cambridge, Massachusetts 02138, USA}
\newcommand{\Houston}{Department of Physics, 
University of Houston, Houston, Texas 77204, USA}
\newcommand{\IHyderabad}{Department of Physics, IIT Hyderabad, Hyderabad, 
502 205, India}
\newcommand{\Hyderabad}{School of Physics, University of Hyderabad, 
Hyderabad, 500 046, India}
\newcommand{\IIT}{Department of Physics,
Illinois Institute of Technology,
Chicago IL 60616, USA}
\newcommand{\Indiana}{Indiana University, Bloomington, Indiana 47405, 
USA}
\newcommand{\INR}{Inst. for Nuclear Research of Russia, Academy of 
Sciences 7a, 60th October Anniversary prospect, Moscow 117312, Russia}
\newcommand{\Iowa}{Department of Physics and Astronomy, Iowa State 
University, Ames, Iowa 50011, USA}
\newcommand{\Irvine}{Department of Physics and Astronomy, 
University of California at Irvine, Irvine, California 92697, USA}
\newcommand{\Jammu}{Department of Physics and Electronics, University of 
Jammu, Jammu Tawi, 180 006, Jammu and Kashmir, India}
\newcommand{\Lebedev}{Nuclear Physics and Astrophysics Division, Lebedev 
Physical 
Institute, Leninsky Prospect 53, 119991 Moscow, Russia}
\newcommand{\MSU}{Department of Physics and Astronomy, Michigan State 
University, East Lansing, Michigan 48824, USA}
\newcommand{\Duluth}{Department of Physics and Astronomy, 
University of Minnesota Duluth, Duluth, Minnesota 55812, USA}
\newcommand{\Minnesota}{School of Physics and Astronomy, University of 
Minnesota Twin Cities, Minneapolis, Minnesota 55455, USA}
\newcommand{\Oxford}{Subdepartment of Particle Physics, 
University of Oxford, Oxford OX1 3RH, United Kingdom}
\newcommand{\Panjab}{Department of Physics, Panjab University, 
Chandigarh, 106 014, India}
\newcommand{\Pitt}{Department of Physics, 
University of Pittsburgh, Pittsburgh, Pennsylvania 15260, USA}
\newcommand{\RAL}{Rutherford Appleton Laboratory, Science and 
Technology Facilities Council, Didcot, OX11 0QX, United Kingdom}
\newcommand{\SAlabama}{Department of Physics, University of 
South Alabama, Mobile, Alabama 36688, USA} 
\newcommand{\Carolina}{Department of Physics and Astronomy, University of 
South Carolina, Columbia, South Carolina 29208, USA}
\newcommand{\SDakota}{South Dakota School of Mines and Technology, Rapid 
City, South Dakota 57701, USA}
\newcommand{\SMU}{Department of Physics, Southern Methodist University, 
Dallas, Texas 75275, USA}
\newcommand{\Stanford}{Department of Physics, Stanford University, 
Stanford, California 94305, USA}
\newcommand{\Sussex}{Department of Physics and Astronomy, University of 
Sussex, Falmer, Brighton BN1 9QH, United Kingdom}
\newcommand{\Syracuse}{Department of Physics, Syracuse University,
Syracuse NY 13210, USA}
\newcommand{\Tennessee}{Department of Physics and Astronomy, 
University of Tennessee, Knoxville, Tennessee 37996, USA}
\newcommand{\Texas}{Department of Physics, University of Texas at Austin, 
Austin, Texas 78712, USA}
\newcommand{\Tufts}{Department of Physics and Astronomy, Tufts University, Medford, 
Massachusetts 02155, USA}
\newcommand{\UCL}{Physics and Astronomy Dept., University College London, 
Gower Street, London WC1E 6BT, United Kingdom}
\newcommand{\Virginia}{Department of Physics, University of Virginia, 
Charlottesville, Virginia 22904, USA}
\newcommand{\WSU}{Department of Mathematics, Statistics, and Physics,
 Wichita State University, 
Wichita, Kansas 67206, USA}
\newcommand{\WandM}{Department of Physics, College of William \& Mary, 
Williamsburg, Virginia 23187, USA}
\newcommand{\Winona}{Department of Physics, Winona State University, P.O. 
Box 5838, Winona, Minnesota 55987, USA}
\newcommand{\Crookston}{Math, Science and Technology Department, University 
of Minnesota -- Crookston, Crookston, Minnesota 56716, USA}
\newcommand{\deceased}{Deceased.}

\affiliation{\ANL}
\affiliation{\Atlantico}
\affiliation{\BHU}
\affiliation{\Caltech}
\affiliation{\Charles}
\affiliation{\Cincinnati}
\affiliation{\Cochin}
\affiliation{\CSU}
\affiliation{\CTU}
\affiliation{\Delhi}
\affiliation{\FNAL}
\affiliation{\UFG}
\affiliation{\Guwahati}
\affiliation{\Harvard}
\affiliation{\Hyderabad}
\affiliation{\IHyderabad}
\affiliation{\Indiana}
\affiliation{\ICS}
\affiliation{\IIT}
\affiliation{\INR}
\affiliation{\IOP}
\affiliation{\Iowa}
\affiliation{\Irvine}
\affiliation{\Jammu}
\affiliation{\JINR}
\affiliation{\Lebedev}
\affiliation{\MSU}
\affiliation{\Duluth}
\affiliation{\Minnesota}
\affiliation{\Panjab}
\affiliation{\SAlabama}
\affiliation{\Carolina}
\affiliation{\SDakota}
\affiliation{\SMU}
\affiliation{\Stanford}
\affiliation{\Sussex}
\affiliation{\Syracuse}
\affiliation{\Tennessee}
\affiliation{\Texas}
\affiliation{\Tufts}
\affiliation{\UCL}
\affiliation{\Virginia}
\affiliation{\WSU}
\affiliation{\WandM}
\affiliation{\Winona}

\author{M.~A.~Acero}
\affiliation{\Atlantico}

\author{P.~Adamson}
\affiliation{\FNAL}


\author{L.~Aliaga}
\affiliation{\FNAL}

\author{T.~Alion}
\affiliation{\Sussex}

\author{V.~Allakhverdian}
\affiliation{\JINR}



\author{N.~Anfimov}
\affiliation{\JINR}


\author{A.~Antoshkin}
\affiliation{\JINR}

\affiliation{\Minnesota}

\author{E.~Arrieta-Diaz}
\affiliation{\SMU}


\author{A.~Aurisano}
\affiliation{\Cincinnati}


\author{A.~Back}
\affiliation{\Iowa}

% ...

\collaboration{The NOvA Collaboration}
\noaffiliation

\date{\today}



\begin{abstract}
We present updated results from the \nova experiment for \numutonumu and \numutonue oscillations from an exposure of $8.85\times10^{20}$ protons on target, which represents an increase of 46\% compared to our previous publication.  The results utilize significant improvements in both the simulations and analysis of the data. A joint fit to the data for \numu disappearance and \nue appearance gives the best fit point as normal mass hierarchy, \dmsq{32} = $2.44\times 10^{-3}\evsq$, \sinsq{23} = 0.56, and \dcp = 1.21$\pi$. The 68.3\% confidence intervals in the normal mass hierarchy are $\dmsq{32} \in [2.37,2.52]\times 10^{-3} \evsq$, $\sinsq{23} \in [0.43,0.51] \cup [0.52,0.60]$, and $\dcp \in [0,0.12\pi] \cup [0.91\pi,2\pi].$    The inverted mass hierarchy is disfavored at 	the 95\% confidence level for all choices of the other oscillation parameters.
\end{abstract}

\pacs{1460.Pq}

\pacs{1460.Pq}
\maketitle


\section{\label{sec:intro} Introduction}

Joint fits of \numutonumu disappearance and \numutonue appearance oscillations in long-baseline neutrino oscillation experiments can provide information on four of the standard neutrino model parameters, $|\dmsq{32}|$, $\theta_{23}$, \dcp, and the mass hierarchy, when augmented by measurements of the other three parameters, \dmsq{21}, $\theta_{12}$, and $\theta_{13}$, from other experiments \cite{pdg}.  Of the four parameters, the first pair are currently most sensitively measured by \numutonumu oscillations and the second pair are most sensitively measured by \numutonue oscillations.  However, the precision with which \numutonue oscillations can measure the second pair of parameters depends on the precision of the measurement of $\theta_{23}$ since that oscillation probability is largely proportional to \sinsqtwo{13}\sinsq{23}.  

The quantity $\tan^2\theta_{23}$ gives the ratio of the coupling of the third neutrino mass state to \numu and \nutau. Whether $\theta_{23} < \pi/4$ (lower octant), $\theta_{23} > \pi/4$ (upper octant), or $\theta_{23} = \pi/4$ (maximal mixing) is important for models and symmetries of neutrino mixing~\cite{ref:Altarelli-Feruglio}.

The determination of the neutrino mass hierarchy is important both for grand unified models~\cite{ref:Altarelli-Feruglio2} and for the interpretation of neutrinoless double beta decay experiments~\cite{ref:Pascoli-Petcov}.  In long-baseline neutrino experiments, it is measured by observing the effect of coherent forward neutrino scattering from electrons in the earth, which enhances \numutonue oscillations for the normal mass hierarchy (NH), $\dmsq{32}>0$, and suppresses them for the inverted mass hierarchy (IH), $\dmsq{32}<0$. For the baselines of current experiments and for fixed baseline length to energy ratio, the magnitude of this effect is approximately proportional to the length of the baseline.  

The amount of CP violation in the lepton sector is proportional to $|\sin\dcp|$.  For \dcp in the range 0 to $2\pi$, \numutonue oscillations are enhanced for $\dcp > \pi$ and suppressed for $\dcp < \pi$, with maximal enhancement at $\dcp = 3\pi/2$ and maximal suppression at $\dcp = \pi/2$.  

In addition to the NOvA results \cite{nova_joint}, previous joint fits of \numutonumu and \numutonue oscillations in long-baseline experiments have been reported by the MINOS \cite{minos} and T2K \cite{t2k} experiments.

The data reported here correspond to the equivalent of $8.85\times10^{20}$ protons on target (POT) in the full \nova Far Detector with a beam line set to focus positively charged mesons, which greatly enhances the neutrino to antineutrino ratio.   This represents a 46\% increase in neutrino flux since our last publication \cite{nova_joint}.
These data were taken between February 6, 2014 and February 20, 2017.

Significant improvements have been made to both the simulations and data analysis. 
The key updates to the simulations include a new data-driven neutrino flux model, an improved treatment of multi-nucleon interactions, and an improved light model including Cherenkov radiation in the scintillator.
The main improvements in the \numu disappearance data analysis are the use of a deep-learning event classifier and the separation of selected events into different samples based on their energy resolution.
The main improvement for the \nue appearance data analysis is the addition of a signal-rich sample that expands the active volume considered.

\section{\label{NOvA} The NO\lowercase{v}A experiment}

NOvA \cite{tdr} is a two-detector, long-baseline neutrino oscillation experiment that samples the Fermilab NuMI neutrino beam \cite{numi} approximately \unit[1]{km} from the source using a Near Detector (ND) and observes the oscillated beam \unit[810]{km} downstream with a Far Detector (FD) near Ash River, MN.
The detectors are functionally identical, scintillating tracker-calorimeters consisting of layered reflective polyvinyl chloride cells filled with a liquid scintillator comprised primarily of mineral oil with a 5\% pseudocumene admixture.
These cells are organized into planes alternating in vertical and horizontal orientation.
The net composition of the detectors is 63\% active material by mass.
Light produced within a cell is collected using a loop of wavelength-shifting optical fiber, which is connected to an avalanche photodiode (APD).

The FD cells are \unit[$3.9 \times 6.6$]{$\text{cm}$} in cross section, with the \unit[6.6]{$\text{cm}$} dimension along the beam direction, and \unit[15.5]{m} long \cite{ref:PVC}.
The FD contains 896 planes, leading to a total mass of \unit[14]{kt}.  
The majority of ND cells are identical to those of the FD apart from being shorter (\unit[3.9]{m} long instead of \unit[15.5]{m}).
To improve muon containment, the downstream end of the ND is a ``muon catcher" composed of a stack of sets of planes in which a pair of one vertically-oriented and one horizontally-oriented scintillator plane is interleaved with one \unit[10]{cm}-thick plane of steel.
There are 11 pairs of scintillator planes separated by 10 steel planes in this sequence.  The vertical planes in this section are \unit[2.6]{m} high.
The ND consists of 214 planes for a total mass of \unit[290]{ton}.

The FD sits \unit[14.6]{mrad} away from the central axis of the NuMI beam.
This off-axis location results in a neutrino flux with a narrow-band energy spectrum centered around \unit[1.9]{GeV} in the FD.
Such a spectrum emphasizes $\numu \rightarrow \nue$ oscillations at this baseline and reduces backgrounds from higher energy neutral current events.
The ND sees a line source and so it receives a much larger spread in off-axis angles than the FD does. The ND is positioned at the same average off-axis angle as the FD to maximize the similarity between the neutrino energy spectrum at its location and that expected at the FD in the absence of oscillations.

The beam is pulsed at an average rate of \unit[0.75]{Hz}.
All of the APD signals above threshold from a large time window around each \unit[10]{$\mu$s} beam spill are retained.
Because the FD is located on the Earth's surface, it is exposed to a substantial cosmic ray flux, which is only partially mitigated by its overburden of \unit[1.2]{m} of concrete plus \unit[15]{cm} of barite.
Therefore, we also use cosmic data taken from \unit[420]{$\mu$s} surrounding the beam spill within beam triggers to obtain a direct measure of the cosmic background in the FD.
Separate periodic minimum-bias triggers of the same length as the beam trigger allow us to collect high-statistics cosmic data for algorithm training and calibration purposes.
As the ND is \unit[100]{m} underground, the cosmic ray flux there is negligible.

\begin{figure}[tb]
  \includegraphics[width=\linewidth]
  {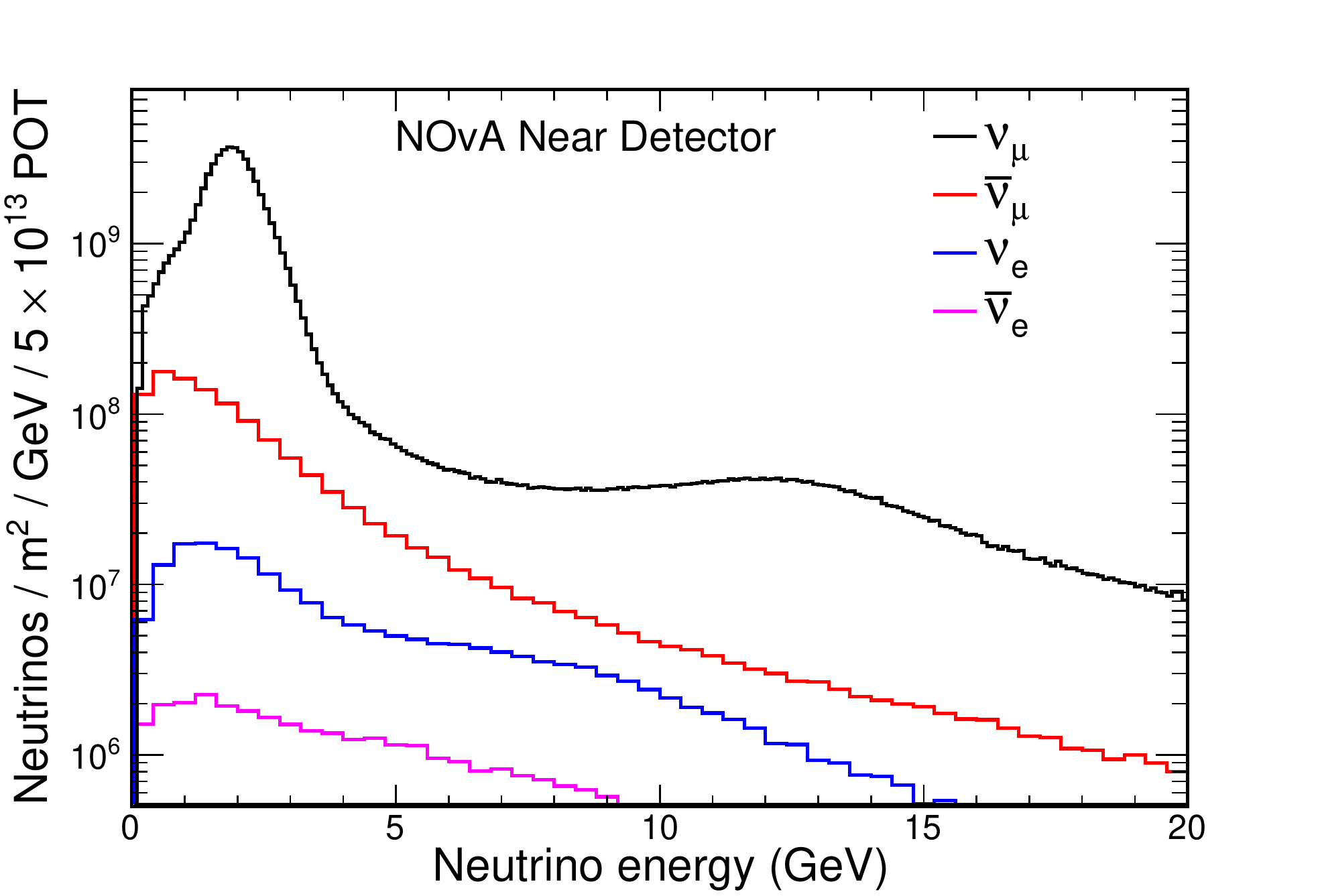}
  \caption{\label{fig:flux}
  Predicted 
  composition of the NuMI beam at the ND with the horns focusing
  positively charged hadrons.  Curves from top to bottom: \numu, $\bar{\nu}_{\mu}$, \nue, $\bar{\nu}_{e}$.  Table~\ref{tab:beam_comp} gives the fractional composition for each neutrino flavor integrated from \unit[1-5]{GeV}.
}

\end{figure}

\section{\label{sim} Simulations}

To assist in calibrating our detectors, determining our analysis criteria, and inferring extracted physical parameters, we rely on predictions generated by a comprehensive simulation suite, which proceeds in stages.
We begin by using \geant/ \cite{geant4} and a detailed model of the beamline geometry to simulate the production of hadrons arising from the collision of the \unit[120]{GeV} primary proton beam with the graphite target \cite{graphite-target}, as well as their subsequent focusing and decay into neutrinos.
The resultant neutrino flux is corrected according to constraints on the hadron spectrum from thin-target hadroproduction data using the \ppfx/ tools developed for the NuMI beam by the MINERvA collaboration \cite{minerva-flux}. The correction applied to the underlying model used in the simulation (FTFP BERT) is in the order of 7-10\% for both, the \numu and \nue flux predictions. The uncertainties are in the order of 8\% in the peak.
Table~\ref{tab:beam_comp} shows simulated predictions of the beam composition at the Near and Far Detectors in the absence of oscillations; the ND predicted spectra from \unit[0-20]{GeV} are shown in Fig.~\ref{fig:flux}.

\begin{table}[h]
  \caption{\label{tab:beam_comp} Predicted beam flux composition in the 1 to 5 GeV neutrino energy region in the absence of oscillations.}
  \begin{tabular}{m{0.35\linewidth}cc} \hline \hline
    Component & \parbox[t]{0.30\linewidth}{ND (\%)} &  \parbox[t]{0.3\linewidth}{FD (\%)} \\ \hline
  \numu            &      93.8     &      94.1 \\
  \numubar         &      5.3      &      4.9 \\
  \nue and \nuebar &      0.9      &      1.0 \\
 \hline \hline
  \end{tabular}
\end{table}

The predicted flux is then used as input to \genie/ \cite{genie-primary,genie-manual}, which simulates neutrino reactions in the variety of materials of which our detectors and their surroundings are composed.
We alter its default interaction model as described below.
Finally, we use a detailed model of our detectors with a combination of \geant/ and custom software to simulate the detectors' photon response to particles outgoing from individual predicted neutrino reactions, including both scintillation and Cherenkov radiation in the active detector materials, as well as the light transport, collection, and digitization processes.
The overall energy scales of both detectors are calibrated using the minimum-ionizing portions of stopping cosmic ray muon tracks.

As in our previous results {\cite{nova_numu,nova_joint,nova_sterile}}, we augment \genie/'s default configuration by enabling its semi-empirical model for Meson Exchange Current (MEC) interactions \cite{katori-empirical-MEC} to account for the likely presence of interactions of neutrinos with nucleon-nucleon pairs in both charged- and neutral-current reactions.
However, in this analysis we no longer reweight the momentum transfer distributions produced by this model, preferring instead to allow fits to the FD data to profile \cite{ref:profile} over the substantially improved systematic uncertainty treatment for this component of the model, as described in Sec.~\ref{systs}. In our central-value prediction we simply increase the rate of MEC interactions by 20\% as suggested by fits to the sample of ND \numuCC candidate events in our ND data.
In addition, we now reweight the output of the default model for quasielastic production to treat the expected effect of long-range nuclear charge screening according to the Random Phase Approximation (RPA) calculations of J. Nieves and collaborators \cite{Valencia-RPA,Gran-RPA}.
Lastly, we continue to reduce the rate of \numuCC nonresonant single pion production with invariant hadronic mass $W < \unit[1.7]{GeV}$ to 41\% of \genie/'s nominal value \cite{nonres-1pi}.

\section{\label{analysis} Data analysis}

In order to infer the oscillation parameters from our data, we compare the spectra observed at the FD with our predictions under various oscillation hypotheses.
This process consists of three steps.
First, we develop selections to retain \nue and \numu charged-current (CC) events and to reject neutral-current (NC) events and cosmogenic activity.
Second, we apply the relevant subset of these selections (excluding, e.g., cosmic rejection criteria) to samples observed at the ND, where both \numu disappearance and \nue appearance are negligible, to constrain our prediction for the selected sample composition.
Finally, we combine the constrained prediction from the previous step with the predicted ratio of the FD and ND spectra, which accounts for geometric differences between the detectors, the beam dispersion, and the effect of oscillations.
The result is used in fits to the neutrino energy spectra of the candidates observed at the FD.
The following sections discuss how this procedure unfolds for each of the two analyses separately.

\subsection{\label{numu} \numu{} disappearance}
\subsubsection{Event selection}

Isolation of samples of candidate events begins with cells whose APD responses are above threshold, known as hits; those neighboring each other in space and time are clustered to produce candidate neutrino events \cite{baird-thesis,slicing}.
We pass hits in event candidates that survive basic quality cuts in timing (relative to the \unit[10]{$\mu$s} beam spill), containment, and contiguity into a deep-learning classifier known as the Convolutional Visual Network (CVN) \cite{cvnpaper}.
CVN applies a series of linear operations, trained over simulated beam and cosmic data event samples, which extract complex, abstract visual features from each event, in a scheme based on techniques from computer vision \cite{szegedy2014googlenet,hinton1986}.
The final step of the classifier is a multilayer perceptron \cite{ref:mlp1,ref:mlp2} that maps the learned features onto a set of normalized classification scores, which range over beam neutrino event hypotheses (\nueCC, \numuCC, \nutauCC, and NC) and a cosmogenic hypothesis.
We retain events whose CVN score for the \numuCC hypothesis exceeds a tuned threshold.

To identify the muon in such events, tracks produced by a Kalman filter algorithm \cite{ref:Kalman,raddatz-thesis,ospanov-thesis} are scored by a $k$-nearest neighbor classifier \cite{ref:kNN} over the following variables: likelihoods in $dE/dx$ and scattering constructed from single-particle hypotheses, total track length, and the fraction of planes along the track consistent with having minimum-ionizing-like $dE/dx$.
The most muon-like of these tracks is taken to be the muon candidate.
Events that have no sufficiently muon-like track are rejected.
We also discard events where any clusters of activity extend to the edges of the detector or where any track besides the muon candidate penetrates into the muon catcher in the ND.
To avoid being considered as cosmogenic, FD events must furthermore be deemed sufficiently signal-like by a boosted decision tree (BDT) \cite{ref:BDT} trained over simulation and cosmic data that considers the positions, directions, and lengths of tracks, as well as the fraction of the event's total hit count associated with the track and the CVN score for the cosmic hypothesis.
According to our simulation, the FD selection efficiency for our basic quality and containment cuts, relative to all true \numuCC events within a fiducial volume, is 41.3\%; the efficiency of the CVN and PID constraints applied to the quality-and-containment sample is 78.1\%.  The final selected sample is 92.7\% \numuCC.
The predicted composition of the sample at various stages in the selection is given in Table~\ref{tab:numu_cutflow}.


\begin{table*}[ht]
  \caption{\label{tab:numu_cutflow} Predicted composition of the \numu CC candidate sample in the FD, in event counts, at various stages in the selection process.  Oscillation parameters used in the prediction are the best fit values from Sec.~\ref{sec:results}.}
\begin{tabular}{m{0.15\linewidth}S[table-format = 3.1]S[table-format = 3.1]SS[table-format = 3.1]Sc}  \hline \hline
Selection     &      
\parbox[t]{0.13\linewidth}{\numutonumu CC}      &      
\parbox[t]{0.13\linewidth}{NC}       &     
\parbox[t]{0.13\linewidth}{ \nueCC }  &     
\parbox[t]{0.13\linewidth} {\nutauCC }        &  
\parbox[t]{0.13\linewidth}{\nuetonumu CC  }      &  
\parbox[t]{0.13\linewidth}  {Cosmic}         \\ \hline \noalign{\vskip 2pt}
No selection    &      963.7    &      612.1    &          126.6    &      9.6      &   0.6      &     $4.91 \times 10^{7} $     \\ 
Containment      &      160.8    &      219.9    &            61.5     &      2.4      &  0.3      &    $1.95 \times 10^{4} $ \\ 
CVN     &      132.1    &      3.0      &           0.3      &      0.4      &   0.2      &    26.4     \\ 
Cosmic BDT     &      126.1    &      2.5      &           0.3      &      0.4      &   0.2      &    5.8      \\ 
 \hline \hline
\end{tabular}
\end{table*}

\subsubsection{Energy estimation and analysis binning}

We reconstruct each event's neutrino energy $E_{\nu}$ using a function of the muon candidate and hadronic remnant energies, which are estimated separately.
The muon candidate energy $E_{\mu}$ is determined from the range of the track, calibrated to true muon energy in our simulation.
We estimate the energy of the hadronic component with a mapping of observed non-muon energy to true non-muon energy also calibrated with the simulation \cite{lein-thesis}.
The resulting neutrino energy resolution over the whole sample is 9.1\% at the FD (11.8\% at the ND due to the lower active fraction of the muon catcher) for \numuCC events.

The precision with which we can measure \sinsqtwo{23} and \dmsq{32} depends on the \numu energy resolution, particularly for events near the disappearance maximum, about \unit[1.6]{GeV} at the NOvA baseline.  Accordingly, we optimize the binning in two ways to get the best effective use of our energy resolution. First, we employ a variable neutrino energy binning with finer bins near the disappearance maximum and coarser bins elsewhere. And, second, 
we further divide the event populations in each energy bin into four populations in reconstructed hadronic energy fraction, $(E_{\nu}-E_{\mu})/E_{\nu}$, which correspond to regions of different neutrino energy resolution \cite{vinton-thesis}.
These divisions are chosen such that the FD populations are of equal size in the unoscillated simulation; however, the boundaries show little sensitivity to the choice of oscillation parameters.
Grouping in this manner has the additional advantage of isolating most background cosmic and beam NC events (those typically mistaken for signal events with energetic hadronic systems) along with events of worst energy resolution into a separate quartile from the three quartiles containing the signal events with better resolution.
The average \numu energy resolution in the FD across the whole energy spectrum is estimated to be  6.2\%, 8.2\%, 10.3\%, and 12.3\% for each quartile, respectively.

\begin{figure}[ht]
  \includegraphics[width=\linewidth]{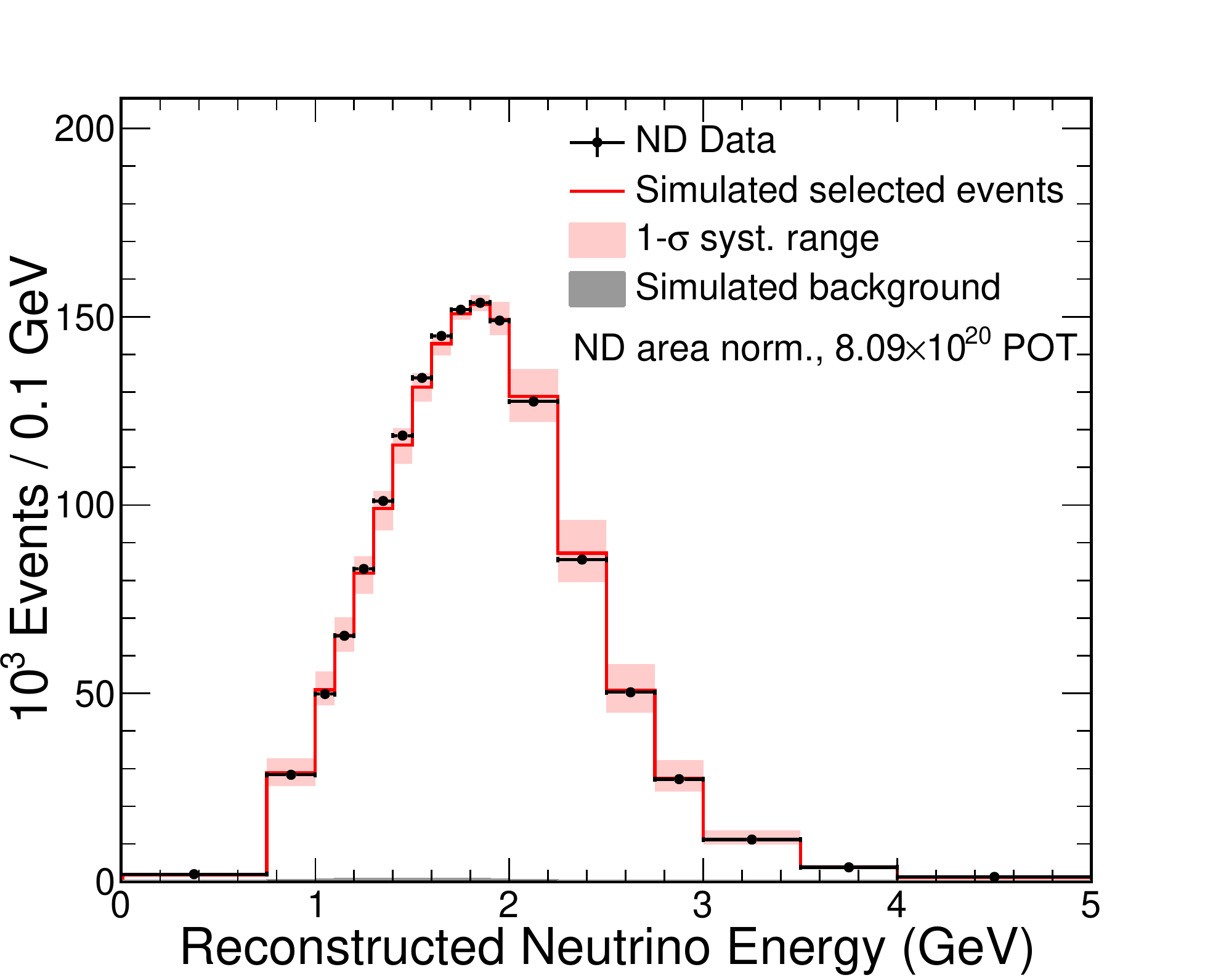}
  \caption{\label{fig:numu_nd}
    Comparison of the reconstructed neutrino energy for selected \numuCC events (black dots) in the ND with area-normalized simulation (red line). Shading represents the bin-to-bin systematic uncertainties.  The gray area, which is nearly indistinguishable from the lower figure boundary, shows the simulated background.}
\end{figure}

\begin{figure*}[ht]
\includegraphics[width=0.8\linewidth]{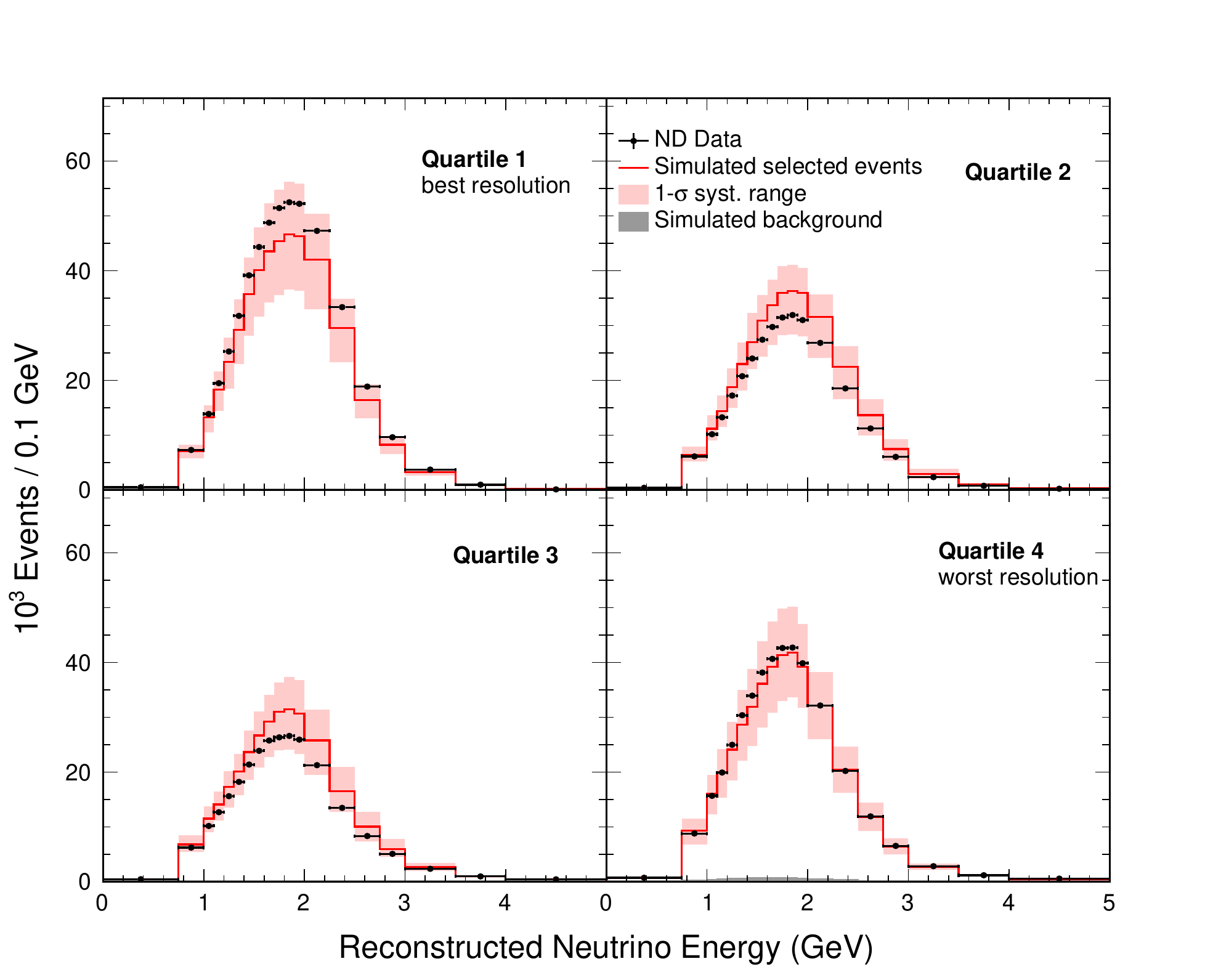}
\caption{\label{fig:numu_nd_quant} Comparison of selected \numuCC candidates (black dots) in the ND data to the prediction (red histograms) in the hadronic energy fraction quartiles, where the prediction is absolutely normalized to the data by exposure.  The expected background contributions (gray) are smaller in the quartiles with better resolution. The shaded band represents the quadrature sum of all systematic uncertainties.  These distributions are the input to the extrapolation procedure described in the text.}
\end{figure*}

Figure~\ref{fig:numu_nd} shows a comparison of the reconstructed neutrino energy for the selected \numuCC events in the ND with simulation shown area-normalized to the data.
The means of the distributions agree to within \unit[10]{MeV} (0.6\%).
Normalizing the prediction by area removes a 1.3\% normalization difference between the data and the simulation and suppresses 10-20\% absolute normalization uncertainties due primarily to our knowledge of the neutrino flux and normalization offsets from cross-section uncertainties.
The remaining uncertainties arise from shape differences.
The full set of uncertainties that are used to compute the error band is described in Sec.~\ref{systs}.
Figure~\ref{fig:numu_nd_quant} shows the corresponding distributions divided into the quartiles.

\subsubsection{\label{subsec:numu extrap} Constraints from the Near Detector data}

As in our previous work \cite{nova_numu}, we obtain a data-driven estimate for the true neutrino energy spectrum using our observed ND data.
To do so, we reweight the simulation in each reconstructed neutrino energy bin to obtain agreement with the ND data, thus correcting the differences observed in Fig.~\ref{fig:numu_nd_quant}.
After subtracting the expected background, which is minimal, we pass the resulting reconstructed neutrino energy spectrum through the migration matrix between reconstructed and true neutrino energies predicted by our ND simulation.
The corrected prediction is then multiplied by the predicted bin-by-bin ratios of the FD and ND true energy spectra, which includes the effects of differing detector geometries and acceptances, beam divergence, and three-flavor oscillations, to obtain an expected FD true energy spectrum.
The latter is finally converted back to reconstructed energy by way of the analogous FD migration matrix.
This constrained signal prediction is summed together with the cosmic prediction, whose reconstructed energy distribution is extracted using events in the minimum-bias trigger passing all the selection criteria and normalized using the \unit[420]{$\mu$s} window around the beam bunch, and a simulation-based beam background prediction to compare to the observed FD data.
In the current analysis, this extrapolation procedure is performed within each hadronic energy fraction range separately so that neutrino reaction types that favor different regions of the elastic-to-inelastic continuum (and thereby have typically different neutrino energy resolution) can be constrained independently.
We find the total number of events in each of the four quartiles, in order from lowest to highest inelasticity, to be adjusted by $+12\%$, $-13\%$, $-13\%$, and $+4\%$ relative to the nominal simulation by this method.

\subsection{\nue appearance}

\subsubsection{Event selection}

We employ the same hit finding and time clustering as in the \numu{} disappearance analysis, and select events whose \nueCC score under the same CVN algorithm exceeds a tuned selection cut.
To further purify the sample of \nueCC candidates, we reconstruct events as follows.
First, we build three-dimensional event vertices using the intersection of lines constructed from Hough transforms applied to each two-dimensional detector view separately \cite{ref:hough,ref:earms}.
Hits in the same view falling roughly along common directions emanating from these vertices are further grouped into ``prongs,'' which are then matched between views based on their extent and energy deposition \cite{ref:fuzzyk,niner-thesis}.
We use these prongs to remove events where the energy of the event is distributed largely transverse to the neutrino beam direction; our simulation and our large sample of cosmic data taken from minimum-bias triggers indicate such events are typically cosmogenic.
We further reject events where the prongs fail containment criteria, where extremely long tracks indicate obvious muons, where there are too many hits for proper reconstruction, or where another event in close proximity in both time and space approaches the top of the detector.
To combat background events from cosmogenic photon showers entering through the back of the detector, where the overburden is thinner, we also cut events which appear to be pointing toward Fermilab rather than away from it.  These events are distinguished by having the number of  planes without hits in the portion of the event closest to Fermilab exceeding the number in the portion farthest from Fermilab, the reverse of the expectation for an electromagnetic shower coming from the neutrino beam direction.
Events surviving these selections form our ``core'' sample in both detectors.
The predicted composition of the FD sample at various stages in this selection is given in Table~\ref{tab:nue_core_cutflow}.

\begin{figure*}[htb]
  \subfloat{%
	\includegraphics[height=0.35\textwidth]{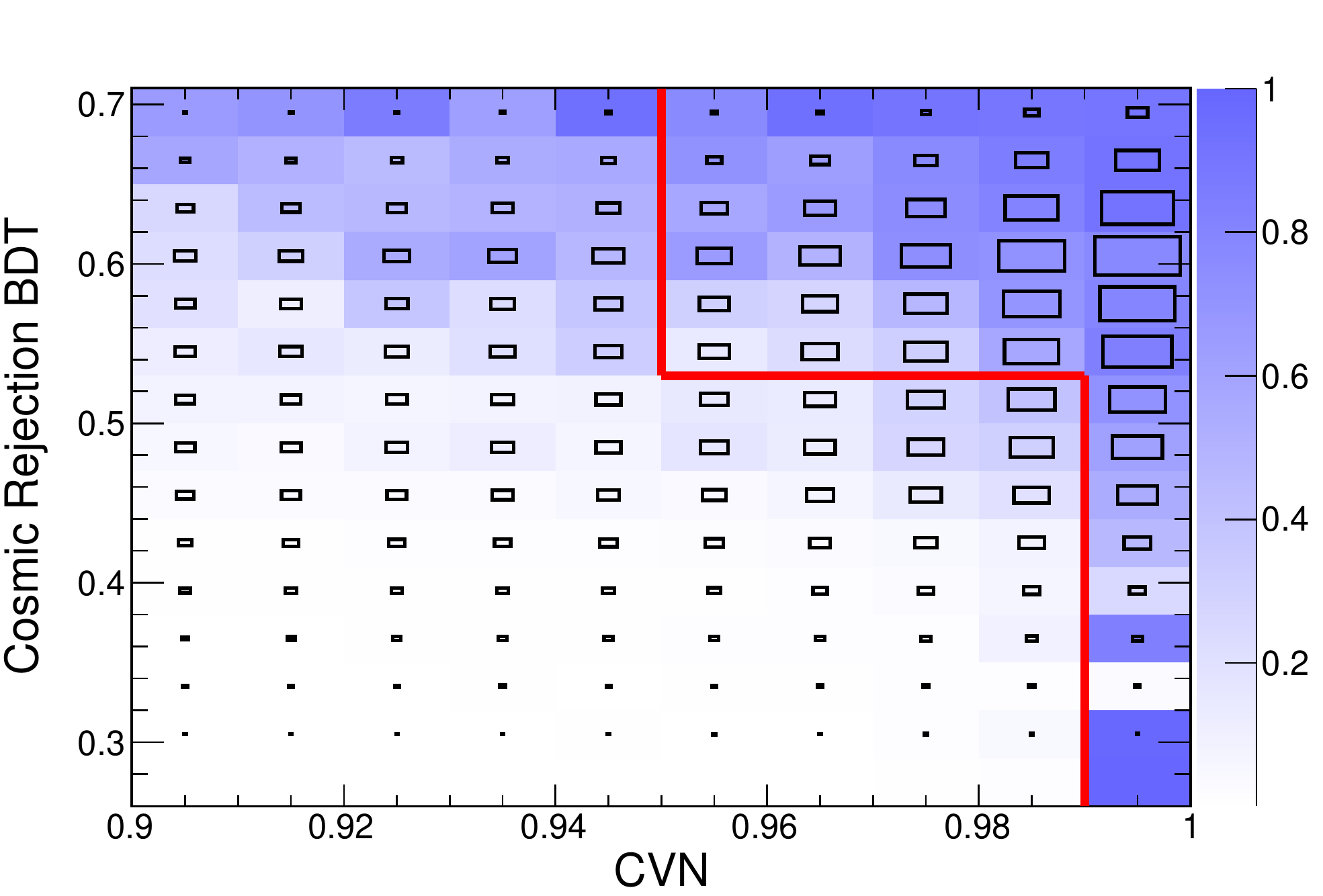}
  }
  \subfloat{%
    \includegraphics[width=0.45\textwidth]{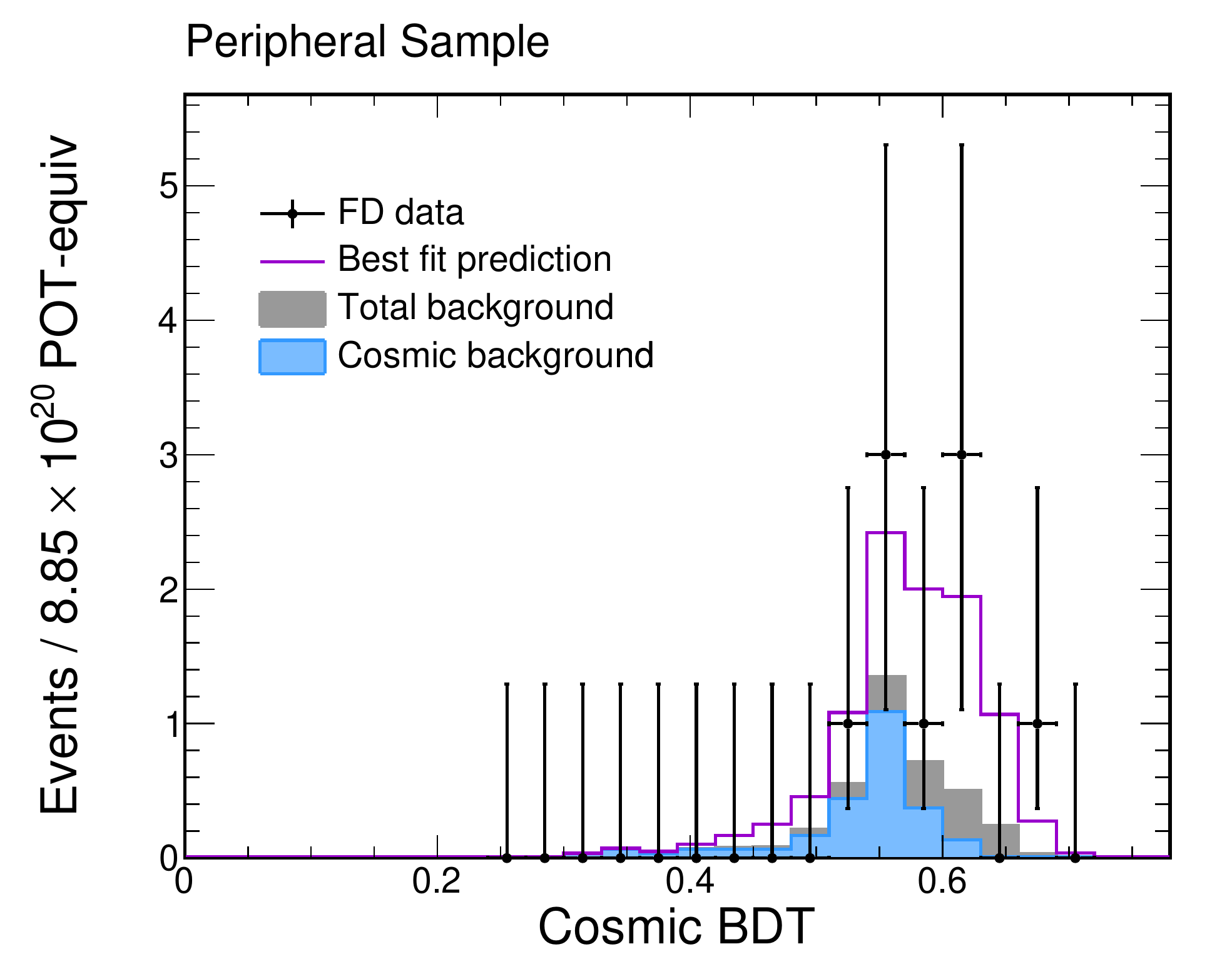}
  }
  \caption{\label{fig:nue_bdt} The peripheral sample is a signal-rich subset of \nue FD candidates  that failed the core cosmic rejection or containment criteria (see text).
  Left: The two-dimensional BDT-CVN space used in the definition of the peripheral sample.  The predicted distribution of \nue appearance signal events (boxes) is shown superimposed on the predicted purity in each bin (shaded color).  The peripheral sample boundary is chosen at the red line: the majority of signal events lie above and to the right and the sample has little cosmogenic contamination there, while events to the left and below are predominantly cosmogenic and are rejected.
  Right: comparison of the observed distribution (black points) of the BDT variable for peripheral events with the prediction (stacked histogram).}
\end{figure*}

\begin{table*}[!ht]
  \caption{\label{tab:nue_core_cutflow} Predicted composition of the core \nue CC candidate sample at the FD, in event counts, at various stages in the selection process.  Oscillation parameters used in the prediction are the best fit values from Sec.~\ref{sec:results}.  These figures do not include the effect of the extrapolation procedure described in Sec.~\ref{subsec:nue ND}.}
  \begin{tabular}{m{0.21\linewidth}SSSSc} \hline \hline
    Selection & \parbox[t]{0.14\linewidth}{$\numu \rightarrow \nue$ CC} &  \parbox[t]{0.14\linewidth}{Beam \nue CC}  &  \parbox[t]{0.14\linewidth}{NC}  &  \parbox[t]{0.14\linewidth}{\numu, \nutau CC}  &  \parbox[t]{0.14\linewidth}{Cosmic}\\ \hline\noalign{\vskip 2pt}
  No selection     &   77.9  &  48.7  &  612.1  &  973.8  &  $4.91 \times 10^{7} $ \\
  Containment/energy cut      &   52.3  &  8.0   &  121.4  & 49.3    &  $2.05 \times 10^{4} $  \\
  Pre-CVN cosmic rejection &   51.3  &  7.9   &  114.3  & 47.0    &  $1.58\times 10^{4} $ \\
  CVN              &   41.4  &  6.0   &    5.3  &  1.3    &  2.0 \\
 \hline \hline
  \end{tabular}
\end{table*}

\begin{table*}[!htb]
  \caption{\label{tab:nue_periph_cutflow} Predicted composition of the peripheral \nue CC candidate sample, in event counts, at two stages in the selection process.  Here ``basic quality'' refers to events that pass beam and detector data quality cuts but fail the core sample containment criteria.  Parameters are as in Table~\ref{tab:nue_core_cutflow}.}
  \begin{tabular}{m{0.21\linewidth}SSSSc} \hline \hline
    Selection & \parbox[t]{0.14\linewidth}{$\numu \rightarrow \nue$ CC} &  \parbox[t]{0.14\linewidth}{Beam \nue CC}  &  \parbox[t]{0.14\linewidth}{NC}  &  \parbox[t]{0.14\linewidth}{\numu, \nutau CC}  &  \parbox[t]{0.14\linewidth}{Cosmic}\\ \hline\noalign{\vskip 2pt}
  Basic quality &   20.4  &  6.6   &  199.9  &  160.9  &  $2.79 \times 10^{6} $  \\
  CVN + BDT          &   5.9   &  1.0   &  0.2    &  0.1    &  2.2 \\
 \hline \hline
  \end{tabular}
\end{table*}

We also construct a second, ``peripheral'' sample of FD events by considering events that have high scores for the CVN \nue hypothesis but which fail the cosmic rejection or containment criteria.
These are subjected to a more focused BDT (distinct from the one mentioned in Sec.~\ref{numu}) trained over the variables used for the containment and cosmic rejection cuts.
The containment variables include the closest distance to the top of the detector and the closest distance to any other face of the detector.
Variables distinguishing cosmogenic from beam-induced activity include the transverse momentum fraction of the event and the number of hits in the event.
Simulation and our cosmic data sample indicate that events in the signal-like regions of both this BDT and CVN are likely to be signal and not the result of externally entering activity and are therefore retained.
Distributions for the peripheral sample illustrating the predicted beam and cosmic response in this BDT and the CVN \nue score, as well as comparing the BDT distribution in data and simulation, are given in Fig.~\ref{fig:nue_bdt}.
Because events on the periphery of the detector are not guaranteed to be fully contained, peripheral events are summed together into a single bin instead of dividing them by the neutrino energy estimate as is done for the core sample.
The FD event counts at two stages of the peripheral selection are noted in Table~\ref{tab:nue_periph_cutflow}.

The ND event sample is predicted to consist of 42\% beam \nue, 30\% NC background, and 28\% \numuCC background.  
These predictions include the effect of the data-driven constraints described in Sec.~\ref{subsec:nue ND}.
The simulated FD efficiency for the basic quality and containment cuts used in the combined core and peripheral selections relative to all true \nue CC events within a fiducial volume is 92.6\%.
The remaining core selections, i.e., CVN and cosmic rejection, retain 58.8\% of the true \nue CC events in the quality-and-containment population.
With the addition of the peripheral sample under the combined CVN+BDT criteria, this figure rises to 67.4\%. 
Improvements to the selection criteria generate an increase of 6.8\% in effective exposure \cite{eff-exposure} relative to our previous results, while the efficiency gain due to the addition of the peripheral sample yields a further increase of 17.4\%.

\subsubsection{Energy estimation and binning}
To estimate the neutrino energy in \nue candidate events, we construct a second-order polynomial in two variables: the sum of the calibrated hit energies from prongs identified as electromagnetic activity and the sum of the energies of hits in the event not within those prongs.
The coefficients of this polynomial are fit to minimize the predicted neutrino energy residuals in selected simulated \nueCC events.
Whether a prong is considered electromagnetic or not is determined by a deep learning single particle classifier that utilizes both information from the prong itself and the full event \cite{psihas-thesis}.
This results in an estimator with 11\% resolution for both appearance signal and beam background \nueCC events in both detectors.

The expected appearance signal has a narrow peak at the \numu disappearance maximum, about \unit[1.6]{GeV}.
Additionally, in this analysis, NC and cosmogenic backgrounds concentrate at low reconstructed energies, and beam \nue backgrounds dominate at high energies.
Based on these considerations, figure-of-merit calculations based on simulation suggest we limit the neutrino energies we consider to be between \unit[1 and 4]{GeV} for the FD core sample and \unit[1-4.5]{GeV} for the peripheral sample.
The corresponding core or peripheral range is used for the ND sample when applying the data constraint detailed in Sec.~\ref{subsec:nue ND}.
Each of these is further subdivided into three ranges in the CVN classifier output so as to concentrate the sample of highest purity together.
The peripheral event sample is treated as a fourth bin.

\subsubsection{\label{subsec:nue ND} Near Detector data constraints}

The procedure for using the ND data in the \nue analysis is similar to that used for \numu, extended to account for the particular natures of the signal and beam background components.
Appeared electron neutrinos arise from oscillated beam muon neutrinos, so the \numu-selected candidates in the ND are used to correct the expected \nue appearance signal with the same procedure detailed in Sec.~\ref{subsec:numu extrap}.
Additionally, the \numu-selected events are used to verify the \nue selection efficiency.
From the \numu data and simulated samples, we create two subsets where the reconstructed muon track is replaced by a simulated electron shower with the same energy and direction \cite{sachdev-thesis}.
The \nue selection criteria are applied to these electron-inserted samples, and the efficiencies for identifying neutrino events in data and simulation, relative to a loose preselection, are found to match within 2\%.

\begin{figure}[htb]
  \includegraphics[width=\linewidth]{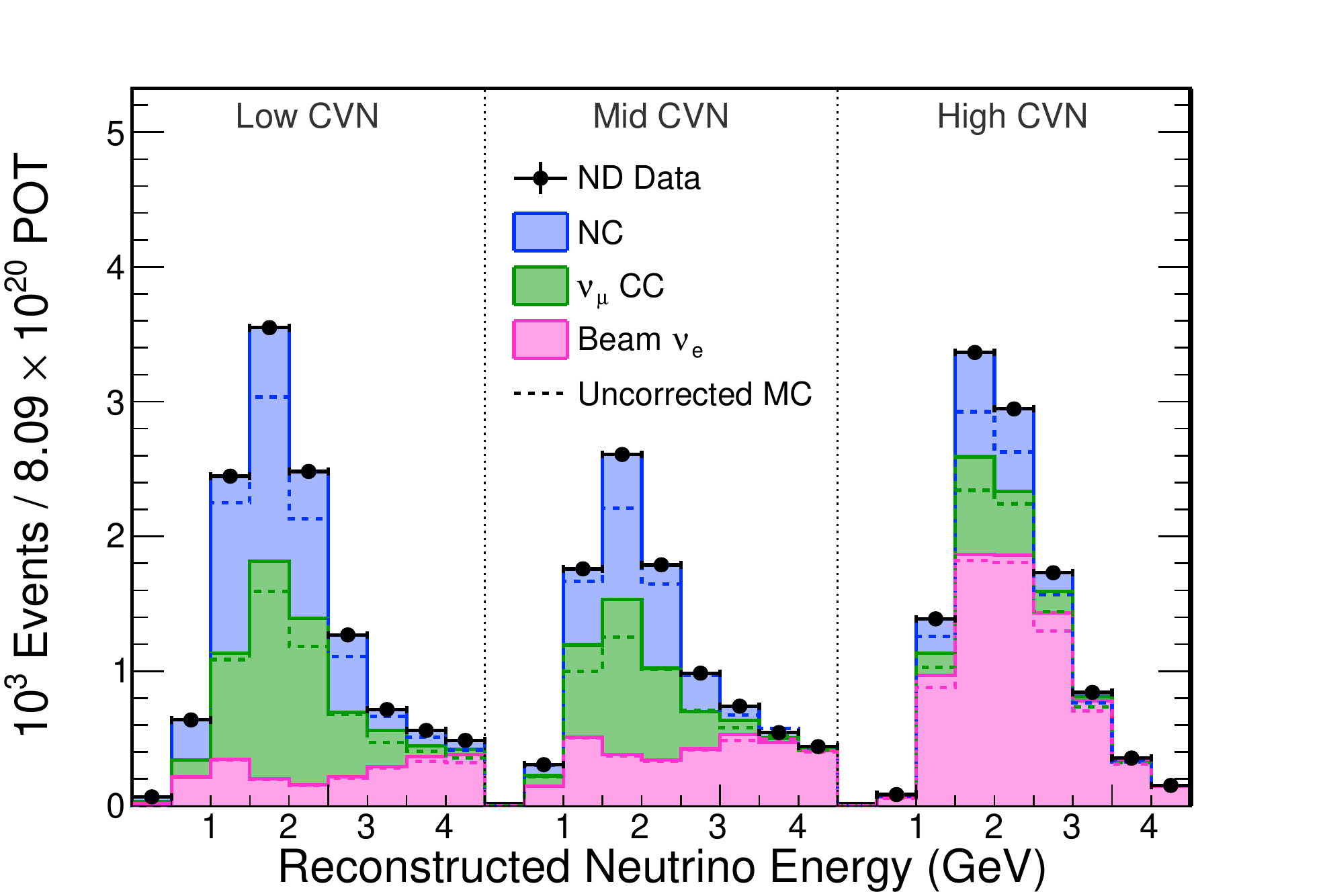}
  \caption{\label{fig:nue_decomp}  The effect of the decomposition and constraint procedure on the predicted ND candidate \nue spectrum; the stacked histogram shows corrected backgrounds (from bottom, beam \nue, \numuCC, NC).  The three panels show the results for each of the CVN classifier bins, ranging left to right from lower to higher purity.  Predictions for each background class prior to correction are given by the dashed lines.  The overall corrections to the normalizations of the yields by category are: beam \nueCC, $+3.0\%$; NC, $+17.0\%$; and \numuCC, $+18.9\%$.}
\end{figure}

As there is no signal and cosmogenic activity is negligible at the ND, the \nueCC candidates at the ND consist entirely of beam background events, originating from CC reactions of the intrinsic \nue component in the beam and mis-identified NC and \numuCC events.
As in our last result \cite{nova_joint}, we use a combination of data-driven methods to ``decompose'' the \nue-selected data into these three categories and constrain them independently.
We examine low- and high-energy \numuCC samples at the ND in order to adjust the yields of the parent hadrons that decay into both \nue and \numu, which constrains the \nue beam background.
We also use the observed distributions of time-delayed electrons from stopping $\mu$ decay in each analysis bin to constrain the ratio of \numuCC and NC interactions.
The resulting decomposition of the selected \nue candidate sample at the ND therefore agrees with the data distribution by construction.
The nominal and constrained predictions are shown compared to the data distribution in Fig.~\ref{fig:nue_decomp}.

The corrections to the beam \nue, NC and \numuCC components are extrapolated to the FD core sample using the bin-by-bin ratios of the FD and ND reconstructed energy spectra, for each of the three CVN ranges.
The predicted beam backgrounds in the FD peripheral sample are corrected according to the results of the extrapolation for the highest CVN bin in the core sample (see Fig. \ref{fig:nue_decomp}).
The sum of the final beam-induced background prediction and the extrapolated signal for given oscillation parameters is added to the measured cosmic-induced backgrounds to compare to the observed FD data.

\section{\label{systs} Systematic uncertainties}

We evaluate the effect of potential systematic uncertainties on our results by reweighting or generating new simulated event samples for each source of uncertainty and repeating the entire measurement, including the extraction of signal and background yields, the computation of migration matrices, and the calculation of the ratios of FD to ND expectations using each modified simulation sample and applying our constraint procedures.

The effect of each of these uncertainties
on the predicted yields of selected \nueCC candidate events is contained in 
Table~\ref{tab:nue_syst}.
We estimate the effects on the extracted oscillation parameters \sinsq{23}, \dmsq{32} and \dcp in the joint fit to be as given in Table~\ref{tab:systs_param}.
These are negligibly different from a \numu-only fit.

\begin{table}[h]
  \caption{\label{tab:nue_syst}  Effect of $1\sigma$ variations of the
    systematic uncertainties on the total \nue
    signal and background predictions.
    Simulated data were used and oscillated with $\dmsq{32}=2.445 \times 10 ^{-3} \evsq$ (NH),
    \sinsq{23}= 0.558, \dcp =  1.21$\pi$. }
  \begin{tabular}{m{0.47\linewidth}SS} \hline \hline
    Source of uncertainty &
    \multicolumn{1}{c}{\parbox[t]{0.20\linewidth}{\nue signal (\%)}} &
    \multicolumn{1}{c}{\parbox[t]{0.27\linewidth}{Total beam \\background (\%)}} \\ \hline
    Cross sections and FSI       &      7.7      &      8.6 \\
    Normalization    &      3.5      &      3.4 \\
    Calibration      &      3.2      &      4.3 \\
	    Detector response        &      0.67     &      2.8 \\
    Neutrino flux    &      0.63     &      0.43 \\
  \nue extrapolation        &      0.36     &      1.2 \\ \hline 
  \mbox{Total systematic} uncertainty &  \multicolumn{1}{c}{9.2} &  \multicolumn{1}{c}{11} \\
  Statistical uncertainty &  \multicolumn{1}{c}{15} &  \multicolumn{1}{c}{22} \\
  \hline 
  Total uncertainty &  \multicolumn{1}{c}{18} &  \multicolumn{1}{c}{25} \\
  \hline \hline
  \end{tabular}
\end{table}

\begin{table}[h]
  \caption{\label{tab:systs_param} Sources of uncertainty and their
    estimated average impact on the oscillation parameters in the
    joint fit. This impact is quantified using the increase in the
    one-dimensional 68\% C.L. interval, relative to the size of the
    interval when only statistical uncertainty is included in the fit.
    Simulated data were used and oscillated with the same parameters
    as in Table~\ref{tab:nue_syst}.
    Given the asymmetry of the
    \sinsq{23} interval with respect to its best fit value, only the
    change in the upper edge is included.
    The total systematic
    uncertainty is calculated by adding the individual components in
    quadrature.}  \resizebox{\linewidth}{!}{ \setlength{\tabcolsep}{1pt}
    \begin{tabular}{
       >{\raggedright}p{0.38\linewidth}
        r
        S[table-align-uncertainty=true,table-number-alignment = left]
        r
        S[table-format = 2.1, table-number-alignment = right]
        c
        S[table-format = 2.1, table-number-alignment = left]
        c
      }
    \hline \hline
      {Source of uncertainty}
      &\multicolumn{2}{c}{\parbox[t]{0.20\linewidth}{Uncertainty \\in \sinsq{23} ($\times 10^{-3}$)}}
      &\multicolumn{4}{c}{\parbox[t]{0.27\linewidth}{Uncertainty \\in \dmsq{32} \small($\times 10^{-6} \evsq$)}}
      &\multicolumn{1}{c}{\parbox[t]{0.20\linewidth}{Uncertainty \\ in \dcp }}
      \\ \hline
            Calibration      &              $+$ &  7.3         &      $+$ & 27& /$-$ &   27        &      $\pm$ 0.05$\pi$\\
      Cross sections and FSI       &  $+$ &  6.9         &      $+$ & 14& /$-$ &    19        &      $\pm$ 0.08$\pi$\\
      Muon energy scale        &      $+$ &  2.4         &      $+$ & 8.5& /$-$ &   12       &      $\pm$ 0.01$\pi$\\
     Normalization    &              $+$ &  4.4         &      $+$ & 7.3& /$-$ &  12       &      $\pm$ 0.05$\pi$\\
     Detector response        &      $+$ &  0.8         &      $+$ & 6.2& /$-$ &   7.7      &      $\pm$ 0.01$\pi$\\
     Neutrino flux    &              $+$ &  1.1         &      $+$ & 4.0& /$-$ &   4.4        &      $\pm$ 0.01$\pi$\\
     \nue extrapolation        &     $+$ &  0.1        &      $+$ & 0.2& /$-$ &   0.7     &      $\pm$ 0.01$\pi$\\ \hline
     {Total systematic uncertainty}     &      $+$ &  12          &      $+$ & 33& /$-$ & 38        &      $\pm$ 0.12$\pi$\\
     Statistical \nohyphens{uncertainty}      &     $+$ &  38          &      $+$ & 75& /$-$ & 84        &      $\pm$ 0.66$\pi$\\ \hline
     Total  uncertainty     &                       $+$ &  40          &      $+$ & 82& /$-$ & 92        &      $\pm$ 0.67$\pi$\\
     \hline \hline
    \end{tabular}
   }
\end{table}

The largest effects on this analysis stem from uncertainty in our calibrations and energy scales, in the cross-section and final-state interaction (FSI) models in \genie/, and in the impact of imperfectly simulated event pileup from the neutrino beam on reconstruction and selection efficiencies at the ND.

\paragraph*{Calibration and energy scale} To evaluate the uncertainty from calibrations and energy scales, which can affect the two detectors differently, we group these uncertainties into absolute (fully positively correlated between detectors) and relative (anticorrelated or uncorrelated) components.
Both absolute and relative muon energy scale uncertainties are $<1\%$ based on a combination of thorough accounting of our detectors' material composition and an examination of the parameters in the Bethe formula for stopping power and the energy-loss model of \geant/.
The overall energy response uncertainty, on the other hand, is driven by uncertainty in our overall calorimetric energy calibration.
To investigate the response, we compare simulated and measured data distributions of numerous channels including the energy deposits of muons originating from cosmogenic- and beam-related activity, the energy spectra of electrons arising from the decay of stopped muons, the invariant mass spectrum of neutral pion decays into photons, and the proton energy scales in ND quasielastic-like events.
The uncertainty we use is guided by the channel exhibiting the largest differences, the proton energy scale, at 5\%.
We take this 5\% uncertainty as both an absolute energy uncertainty, correlated between the two detectors, and a separate 5\% relative uncertainty, since there are not sufficient quasielastic-like events to perform this check at the FD.

\begin{figure*}[!t]

  \includegraphics[width=0.8\textwidth]{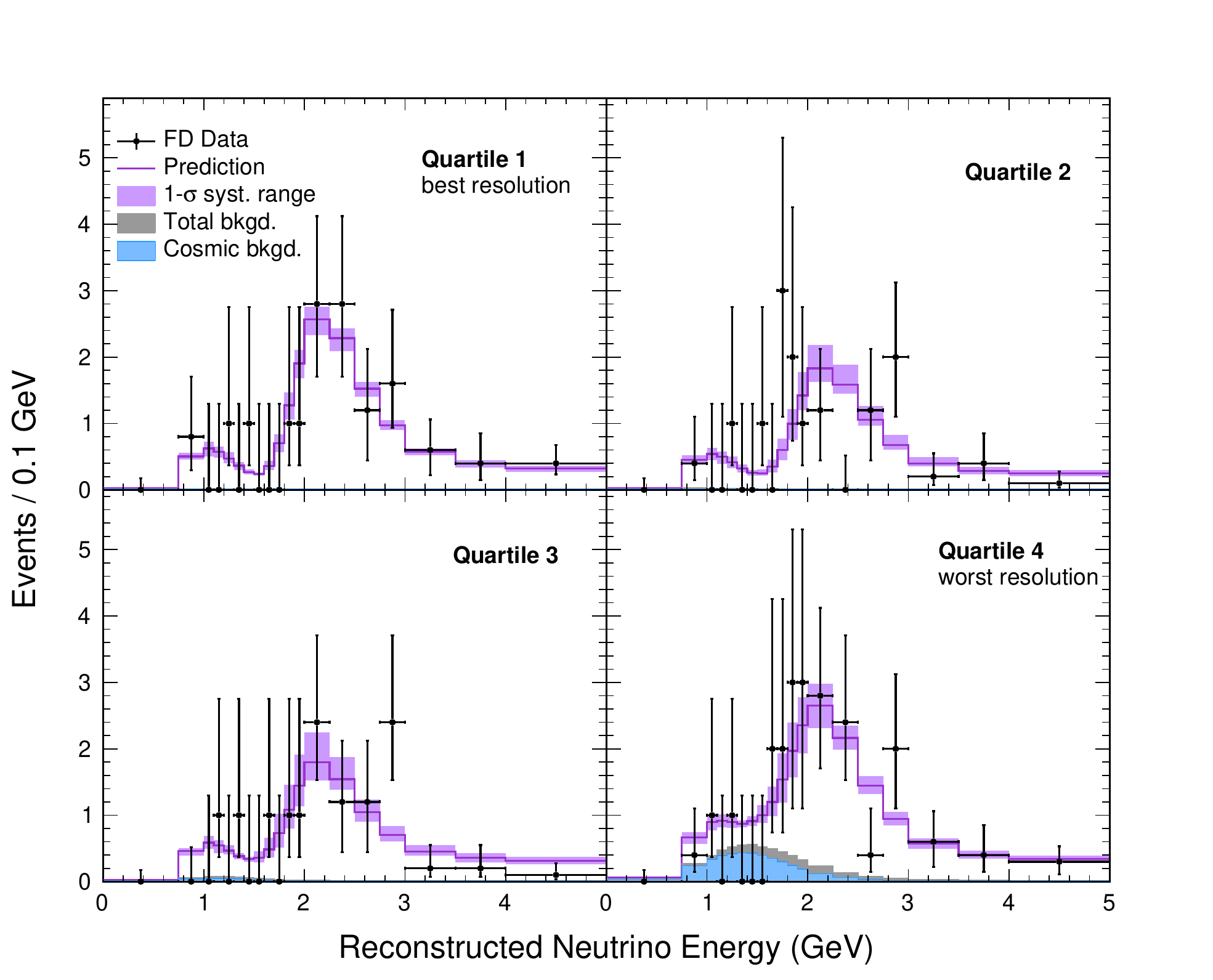}
  \caption{\label{fig:numu_spectr} Comparison of the reconstructed
    energy spectra of selected \numuCC candidates in FD data (black dots) and
    best-fit prediction (purple). The sample is split into four
    reconstructed hadronic energy fraction quartiles labeled 1 through
    4, where 1 (4) has the best (worst) energy resolution. The
    majority of the total background (gray, upper) including the cosmogenic subcomponent
    (blue, lower) lies in the fourth quartile.}
 \end{figure*}

\paragraph*{Cross sections and FSI} Estimates for the majority of the cross section and FSI uncertainties that we consider are obtained using the event reweighting framework in \genie/ \cite{genie-manual}.
However, ongoing effort in the neutrino cross section community and the NOvA ND data suggest some modifications are necessary.
First, we apply additional uncertainty to the energy- and momentum-transfer-dependence of CC quasielastic (CCQE) scattering due to long-range nuclear correlations \cite{Valencia-RPA-unc} according to the prescription in Ref.~\cite{Gran-RPA}.
Second, as the detailed nature of MEC interactions is not well understood, we construct uncertainties for the neutrino energy dependence, energy-transfer  dependence, and final-state nucleon-nucleon pair composition based on a survey of available theoretical treatments \cite{Gran-Valencia-2p2h,Martini-2p2h,SuSA-2p2h}.
The normalization of the MEC component is recomputed under each of these uncertainties using the same fit procedure used to arrive at the 20\% scale factor for the central value prediction.
Third, it is now believed that the inflated value of the axial mass in quasielastic scattering ($M_A^{QE}$) obtained in recent neutrino-nucleus scattering experiments relative to the light liquid bubble chamber measurements is due to nuclear effects that we are now treating explicitly with the foregoing \cite{nu-xsec-review}.
We thus reduce \genie/'s uncertainty for $M_A^{QE}$ to $\pm 5\%$ (a conservative estimate of the bubble chamber range \cite{MAQE-BBBA,MAQE-Zexp}) from its default of ${}^{+25\%}_{-15\%}$, while retaining \genie/'s central value $M_{A}^{QE} = \unit[0.99]{GeV/c^{2}}$.
Fourth, we increase the uncertainty applied to nonresonant pion production with three or more pions and invariant hadronic mass of $W<\unit[3]{GeV}$ to 50\% to match the default for 1- and 2-pion cases, based on data-simulation disagreements observed in the ND data.
Fifth, and finally, we introduce two separate 2\% uncertainties on the ratio of \nueCC and \numuCC cross sections: one to account for potential differences between them due to radiative corrections, and one to consider the possibility of second-class currents in CCQE events \cite{DayMcF-numu-nue-diff,t2k}.

To validate the uncertainties assigned by \genie/ to the NC backgrounds in our analyses, we performed a study within the \numuCC candidate sample in the ND that measured the rates of neutrons that were produced at the ends of tracks and subsequently recaptured, emitting photons. This study was done by investigating time-delayed activity consistent with a neutron capture, taking into account the tail of the Michel electron time spectrum. The neutron rate is different for the mostly $\mu^{-}$ identified in \numuCC reactions versus the mostly $\pi^{\pm}$ in NC.
This study suggested that the NC cross-section uncertainties provided by GENIE, combined together with the calibration uncertainties mentioned previously, account for any differences between data and simulation.
Therefore we no longer include the \textit{ad hoc} 100\% additional uncertainty on NC backgrounds used in previous results \cite{nova_numu,nova_joint}.

\paragraph*{Normalization} We quantify the uncertainty arising from potential imperfections in the simulation of beam-induced pileup in the ND by overlaying a single extra simulated event onto samples of both simulated and data events.
We then examine the selection efficiency of this extra event and assign the 3.5\% difference between the data and simulation samples as a conservative uncertainty on the normalization of the ND rate.  These are added in quadrature with much smaller uncertainties in the detector mass and the total beam exposure to yield an overall normalization systematic.

\paragraph*{Other} Other contributions to our systematic uncertainty budget are associated with the improved \ppfx/ flux prediction and potential differences between the acceptances of the ND \numu selection criteria and the FD \nue sample into which the ND corrections are extrapolated in the \nue analysis.
Also substantially reduced are the uncertainties in the light response model used for detector simulation.
Previous fits of the parameters in the Birks model for scintillator quenching with a second-order term \cite{chou-scint}, using proton tracks in candidate ND \numuCC quasielastic-like events in data, obtained values inconsistent with other measurements of Birks quenching in liquid scintillator \cite{KamLAND-scint,Borexino-scint}.
Previous results therefore used a variation with the other measurements' values to compute an uncertainty.
With the addition of Cherenkov light in scintillator to our detector model, however, we find a best fit at  the same values preferred by other experiments.
To quantify any residual uncertainty in the light model, in this analysis we take alternate predictions where we alter the scintillation and Cherenkov photon yields in the model within the tolerance of agreement with the ND data while holding the muon response fixed (since it is set by our calibration procedure).

\section{\label{sec:results} Results}

We performed a blind analysis in which the FD data were analyzed only after all aspects of the analysis had been specified.
An independent implementation of the methods described in Secs.~\ref{analysis}-\ref{systs} for incorporating the Near Detector data constraint and assessing the impact of systematic uncertainties, as well as extracting oscillation parameters via likelihood fitting, was used to check the analysis presented in this paper.
It produced results consistent with those shown in the following sections.  

\subsection{\numu disappearance data}

After selection, 126 \numuCC candidates are observed in the FD.
In the absence of oscillations, we would have expected $720.3^{+67.4}_{-47.0} \text{ (syst.)}$ \numuCC candidates based on the extrapolation from the Near Detector, including an expected background of 5.8 misidentified cosmic rays and 3.4 misidentified neutrino events of other types.

 \begin{figure}[htb]
	  \includegraphics[width=\linewidth]{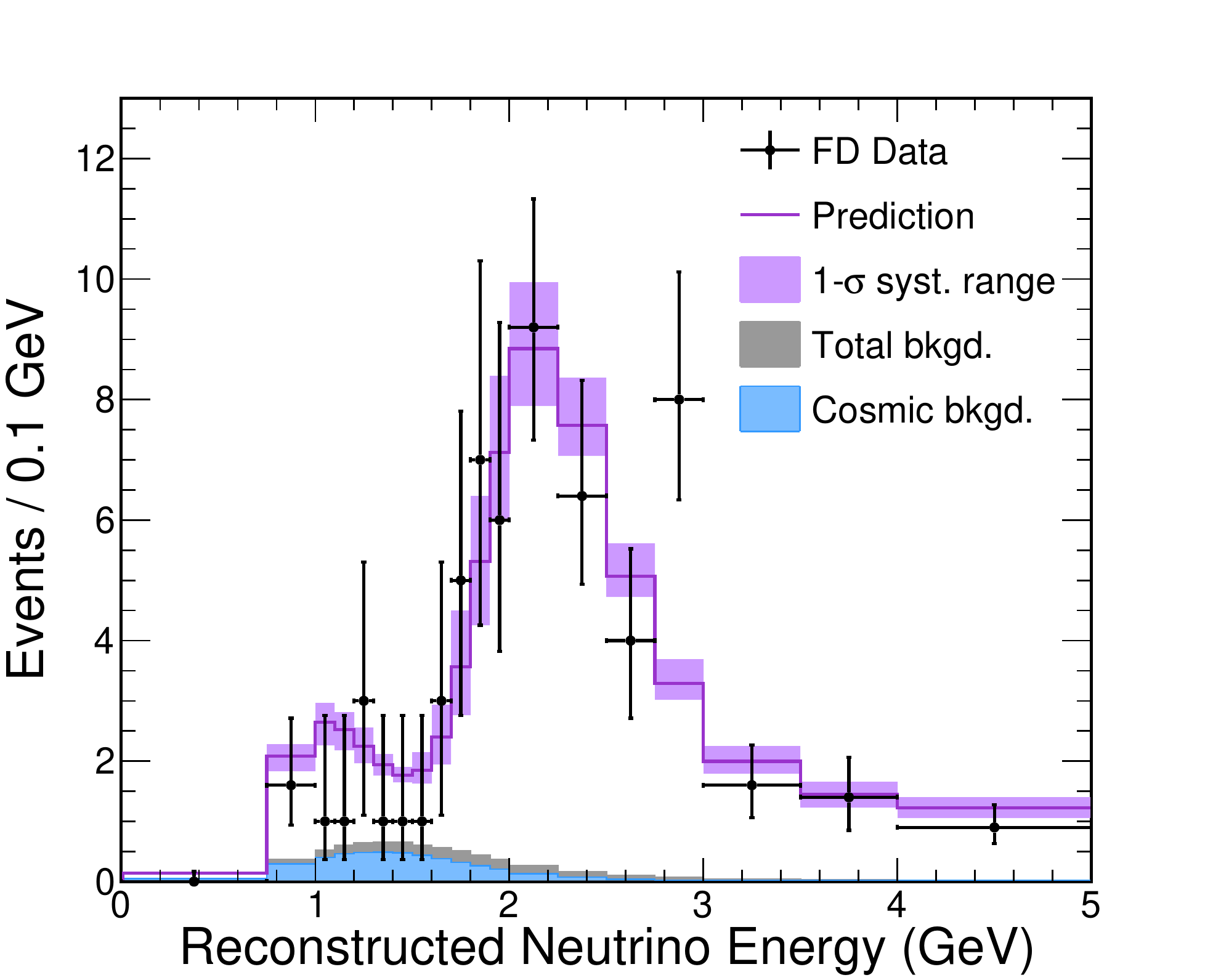}
  \caption{\label{fig:numu_combined} Data from
    Fig.~\ref{fig:numu_spectr} summed over the four quartiles.}
\end{figure}

Figure~\ref{fig:numu_spectr} shows the observed energy spectrum in each quartile and the corresponding best fit predictions. As noted earlier, most of the predicted background appears in the fourth (worst resolution) quartile.  Figure~\ref{fig:numu_combined} shows the data of Fig.~\ref{fig:numu_spectr} summed over all of the quartiles.
The neutrino energy spectrum exhibits a sharp dip at about \unit[1.6]{GeV}.  Essentially, \sinsqtwo{23} corresponds to the depth of the dip and \dmsq{32} corresponds to its location.  Both of these measurements are sensitive to the energy resolution, so we expect the best measurement in the quartile with best energy resolution.

\subsection{\nue appearance data}

After selection we observe 66 \nueCC candidate events in the FD including an expected background of $20.3\pm 2.0 \text{ (syst.)}$ events.  The composition of the expected background is estimated to be 7.3 beam \nueCC events, 6.4 NC events, 1.3 \numuCC events, 0.4 \nutauCC events, and 4.9 cosmic rays.

\begin{figure}[htb]
  \includegraphics[width=\linewidth]{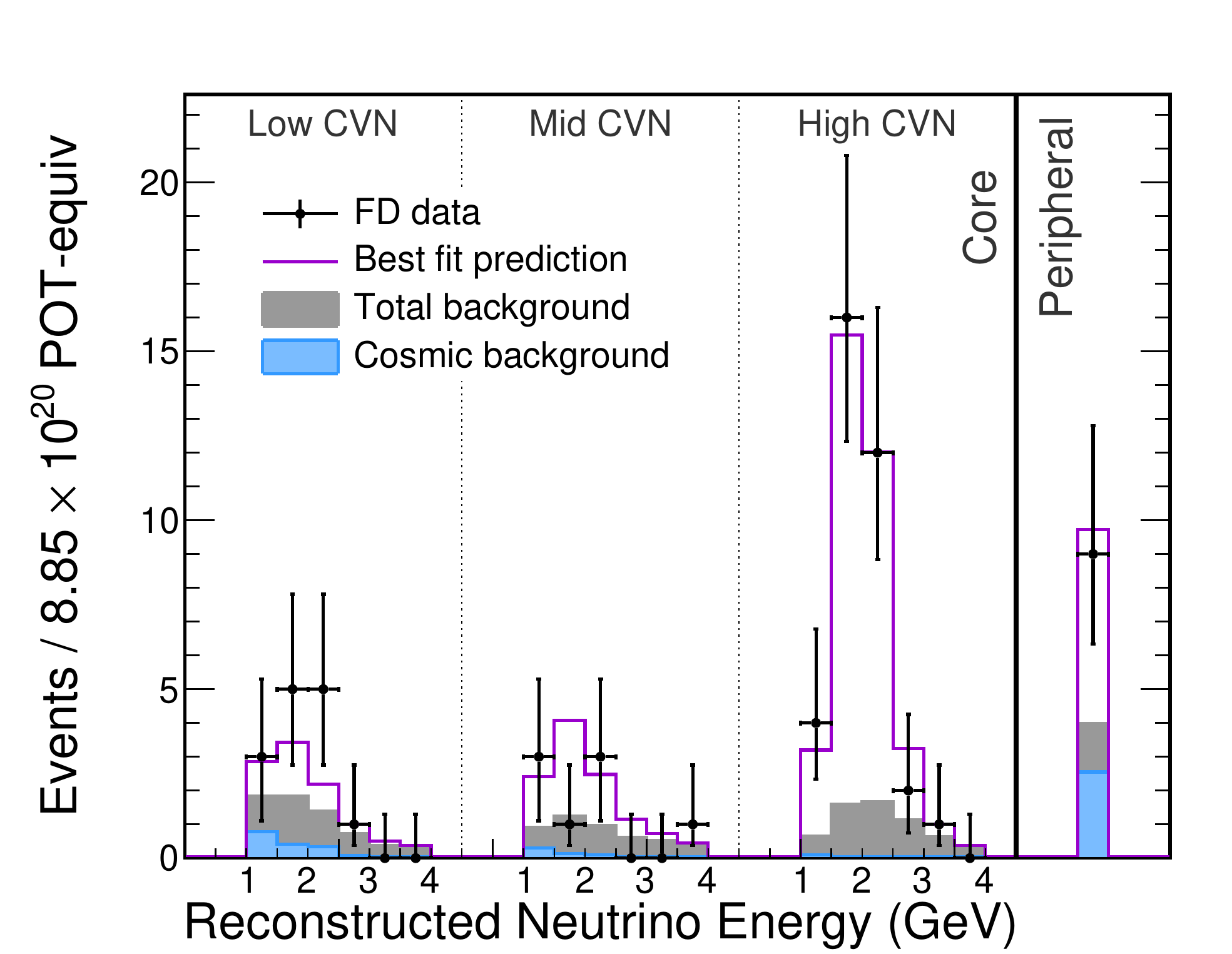}
  \caption{\label{fig:nue_spectr} Comparison of the neutrino energy spectra of selected \nueCC candidates in the FD data (black dots) with the best fit prediction (purple lines) in the three CVN classifier bins and the peripheral sample.  The total expected background (gray, upper) and the cosmic component of it (blue, lower) are shown as shaded areas. 
The events in the peripheral bin have energies between 1 and 4.5 GeV.}
\end{figure}

\begin{figure}[htb]
  \includegraphics[width=\linewidth]{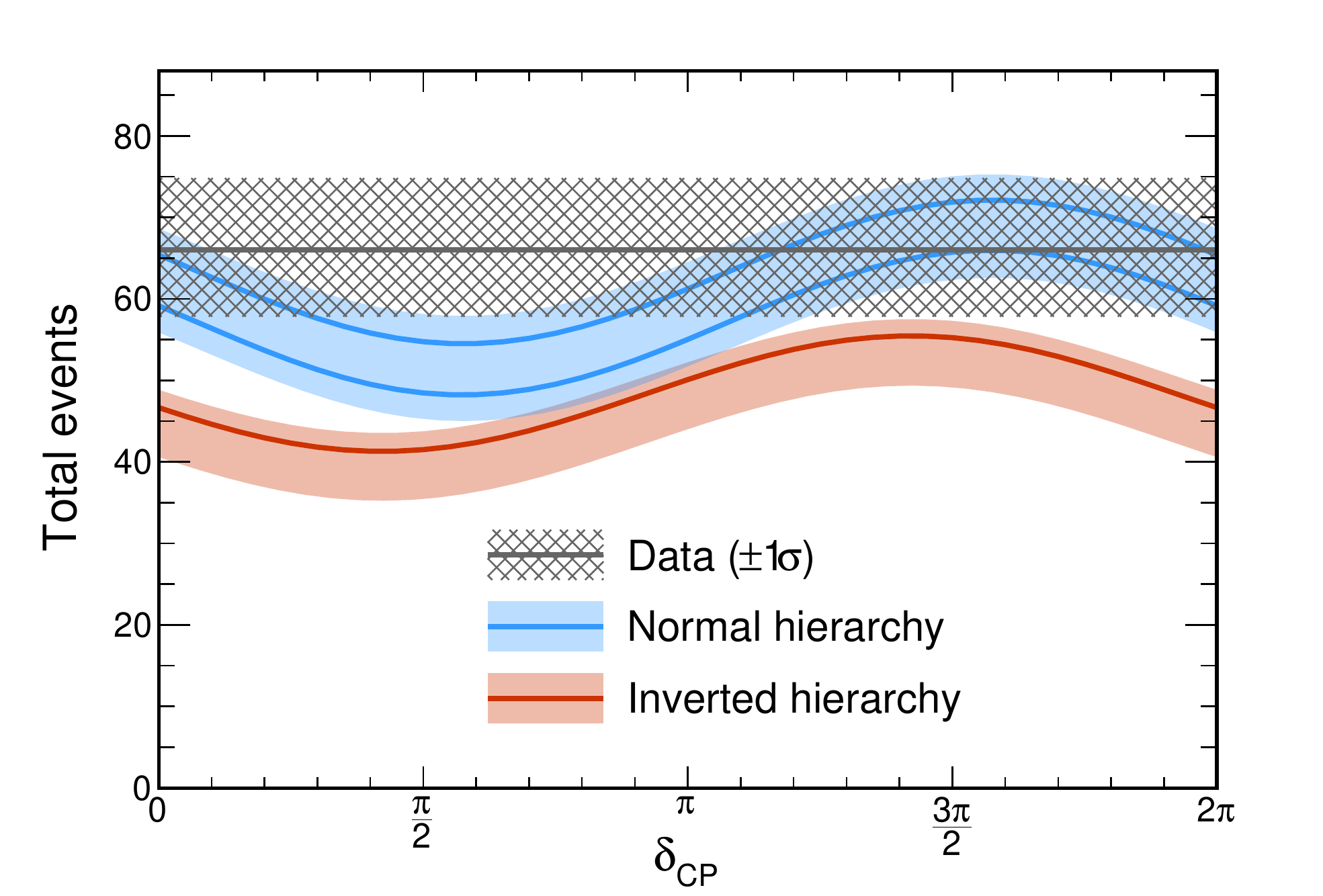}
  \caption{\label{fig:nue_number} 
    Total number of \nueCC candidate events observed in the FD (gray) compared to the prediction (color) as a function of \dcp.
    The color lines correspond to the best  fit values of \sinsq{23} and \dmsq{32} from
    Table~\ref{tab:best_fits}, with
    the upper two curves (blue) representing two octants in the normal mass hierarchy ($\dmsq{32}>0$) and the lower curve (red) the inverted hierarchy ($\dmsq{32}<0$).
    The color bands correspond to $0.43 \leq \sinsq{23} \leq 0.60$. All other parameters are held fixed at the best-fit values.
  }
\end{figure}

Figure~\ref{fig:nue_spectr} shows the distribution of these events as a function of the reconstructed neutrino energy for the three CVN classifier bins and for the peripheral sample, along with the expected background contributions and the best fit predictions.
To give some context to the number of observed \nue events, Fig.~\ref{fig:nue_number} shows the number of events expected for the best fit values of \dmsq{32} and \sinsq{23} as a function of \dcp, for the two possible mass hierarchies.

\subsection{Joint fit results}

We have performed a simultaneous fit to the binned data shown in Figs.~\ref{fig:numu_spectr} and \ref{fig:nue_spectr}.
Systematic uncertainties are incorporated into the fit as nuisance parameters with Gaussian penalty terms.
Where systematic uncertainties are common between the two data sets, the nuisance parameters associated with the effect are correlated appropriately.
In making these fits  and in the contours and significance levels that follow, we used the following values for physics parameters measured by other experiments \cite{pdg}: $\dmsq{21} = (7.53\pm 0.18)\times 10^{-5}\evsq$, $\sin^2\theta_{12} = 0.307\substack{+0.013 \\ -0.012}$, $\sin^2\theta_{13} = 0.0210 \pm 0.0011$.
We use a matter density computed for the average depth of the NuMI beam in the Earth's crust for the NOvA baseline of 810 km using the CRUST2.0 model \cite{ref:crust}, $\rho = \unit[2.84]{\mathrm{g/cm^3}}$.

\subsubsection{Best fits}

\begin{figure}[htb]
\includegraphics[width=\linewidth]{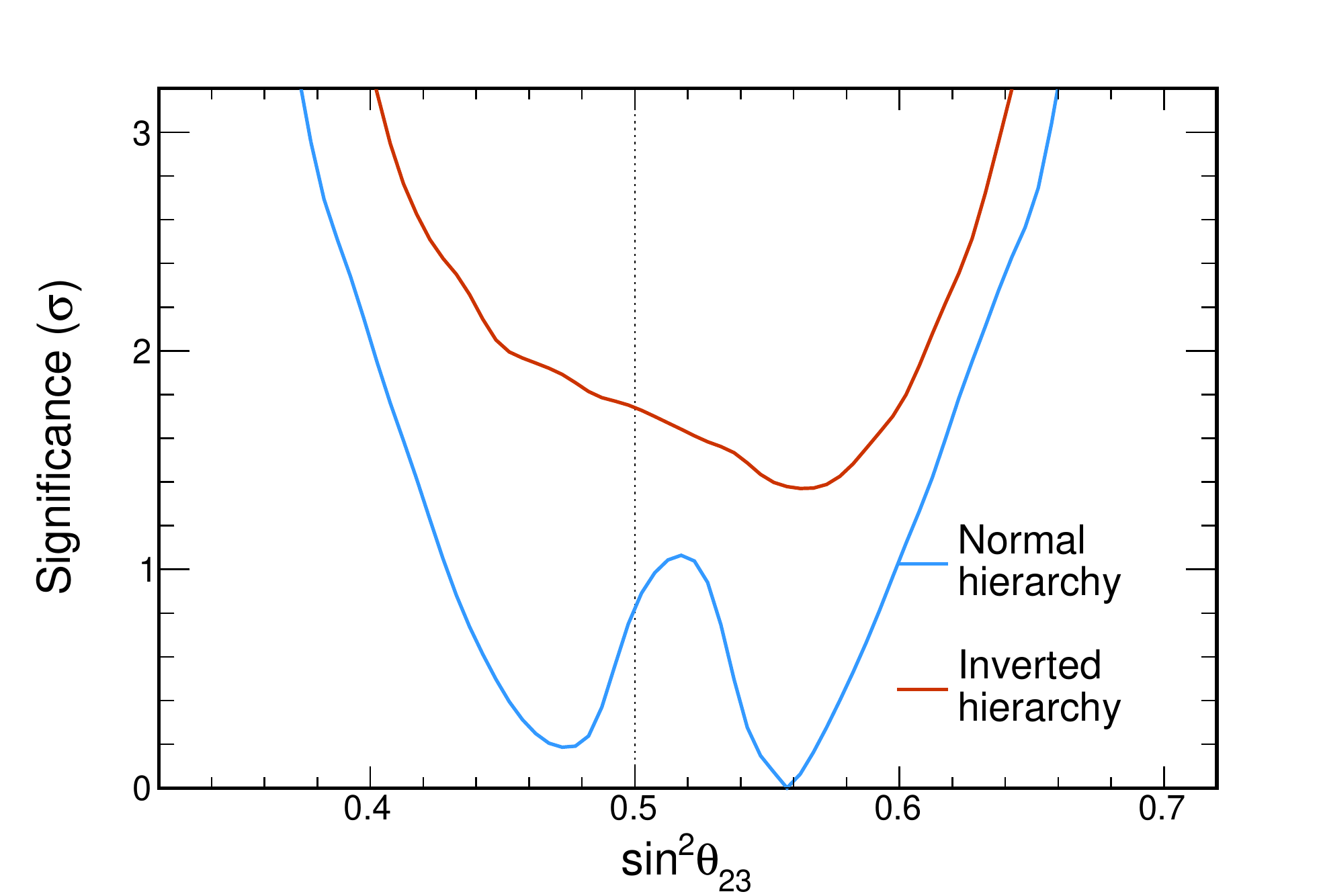}
\caption{\label{fig:th23}
Significance at which each value of \sinsq{23} is disfavored
in the normal (blue, lower) or inverted (red, upper) mass hierarchy.
The vertical dotted line indicates the point of maximal mixing.}
\end{figure}

Table~\ref{tab:best_fits} gives the parameter values at the best fit point in each relevant mass hierarchy and $\theta_{23}$ octant combination.  The top line shows the overall best fit, which occurs in the normal mass hierarchy and the upper $\theta_{23}$ octant; the middle line shows best fit in the lower $\theta_{23}$ octant for the normal mass hierarchy, which is  only slightly less significant; and the bottom line shows the best fit in the inverted mass hierarchy, which is disfavored largely because it predicts fewer \nue appearance events than are observed. The column labeled $\Delta\chi^2$ represents the difference in $\chi^2$ between the fit and the overall best fit, where $\chi^2$ in this case is $-2{\text {ln}}\mathcal{L}$ with $\mathcal{L}$ being the likelihood function calculated using Poisson statistics plus Gaussian penalty terms for the systematic uncertainties. There are no best fit values in the inverted mass hierarchy and lower $\theta_{23}$ octant because the likelihood has no local maximum in this hierarchy-octant region, as will become clear in Fig.~\ref{fig:th23}.  The $\chi^2$ for the overall best fit is 84.6 for 72 degrees of freedom.

The precision measurements of \sinsq{23} and \dmsq{32} come from the \numu disappearance data.  A fit to these data alone gives essentially the same values for these parameters in the normal mass hierarchy.  However, the best joint \numu{}-\nue{} fit pulls the value of $|\dmsq{32}|$ up by $0.04 \times10 ^{-3} \evsq$ from the \numu disappearance only fit in the inverted mass hierarchy.

\begin{table}[h]
  \caption{\label{tab:best_fits} Best fit values. See text for further explanation.}
\resizebox{\linewidth}{!}{
    \begin{tabular}{lcccc}
    \hline \hline
      Hierarchy/Octant 
      & {\dcp ($\pi$)}
      & {\sinsq{23}}
      & \parbox[t]{0.22\linewidth}{\dmsq{32} ($10^{-3} \evsq)$ } &
      {$\Delta\chi^2$}
     \\ \hline
     Normal/Upper    &     1.21       &     0.56     &  \phantom{$-$}2.44 & 0.00\\
     Normal/Lower    &     1.46       &     0.47     &  \phantom{$-$}2.45  & 0.13\\
     Inverted/Upper  &     1.46       &     0.56
     &   $-$2.51  & 2.54 \\

     \hline \hline
   \end{tabular}
   }
\end{table}

\begin{figure}[!htb]
  \includegraphics[width=\linewidth,trim=0 45 0 0,clip]{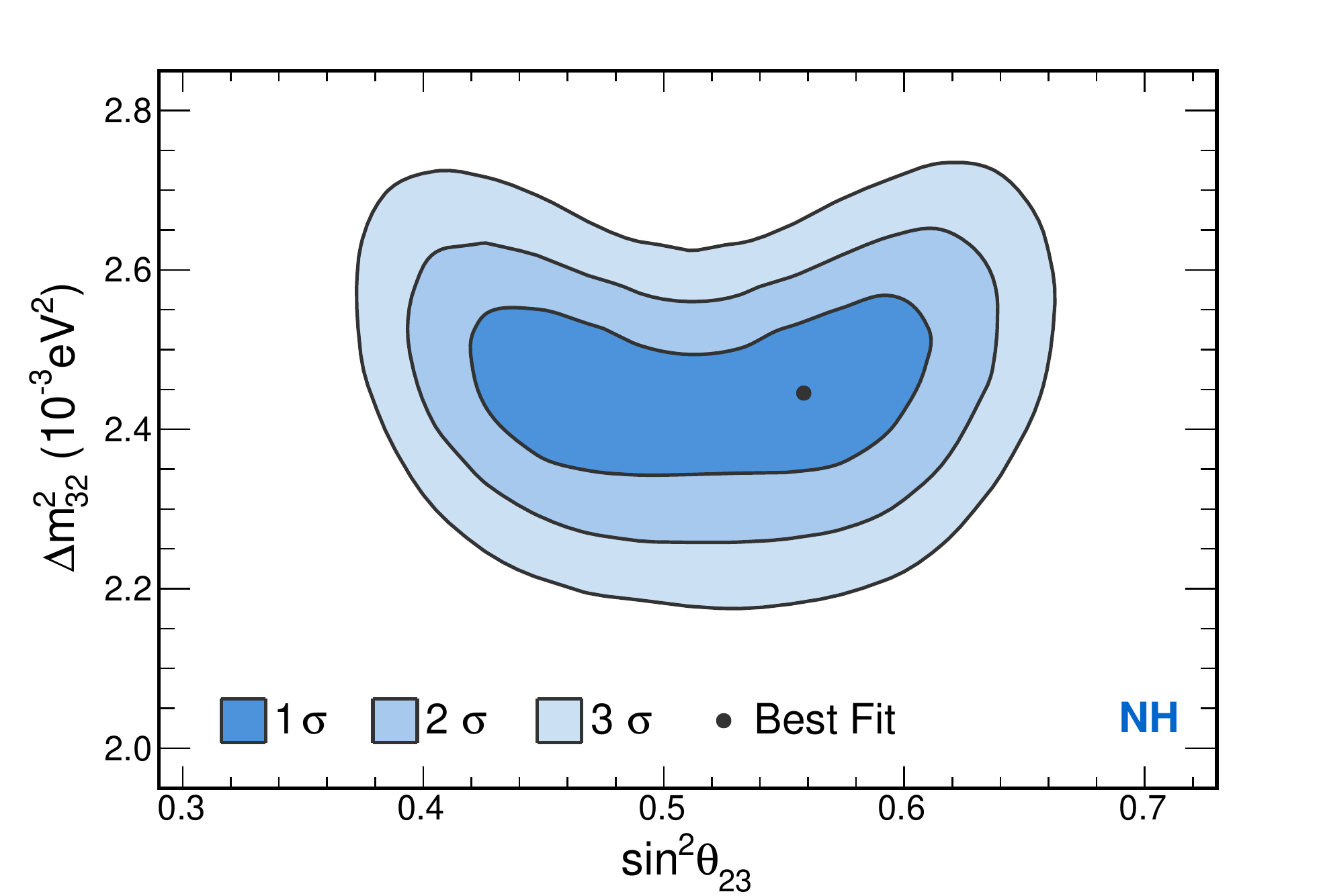}
  \vspace{-.2em}
  \includegraphics[width=\linewidth,trim=0 0 0 25,clip]{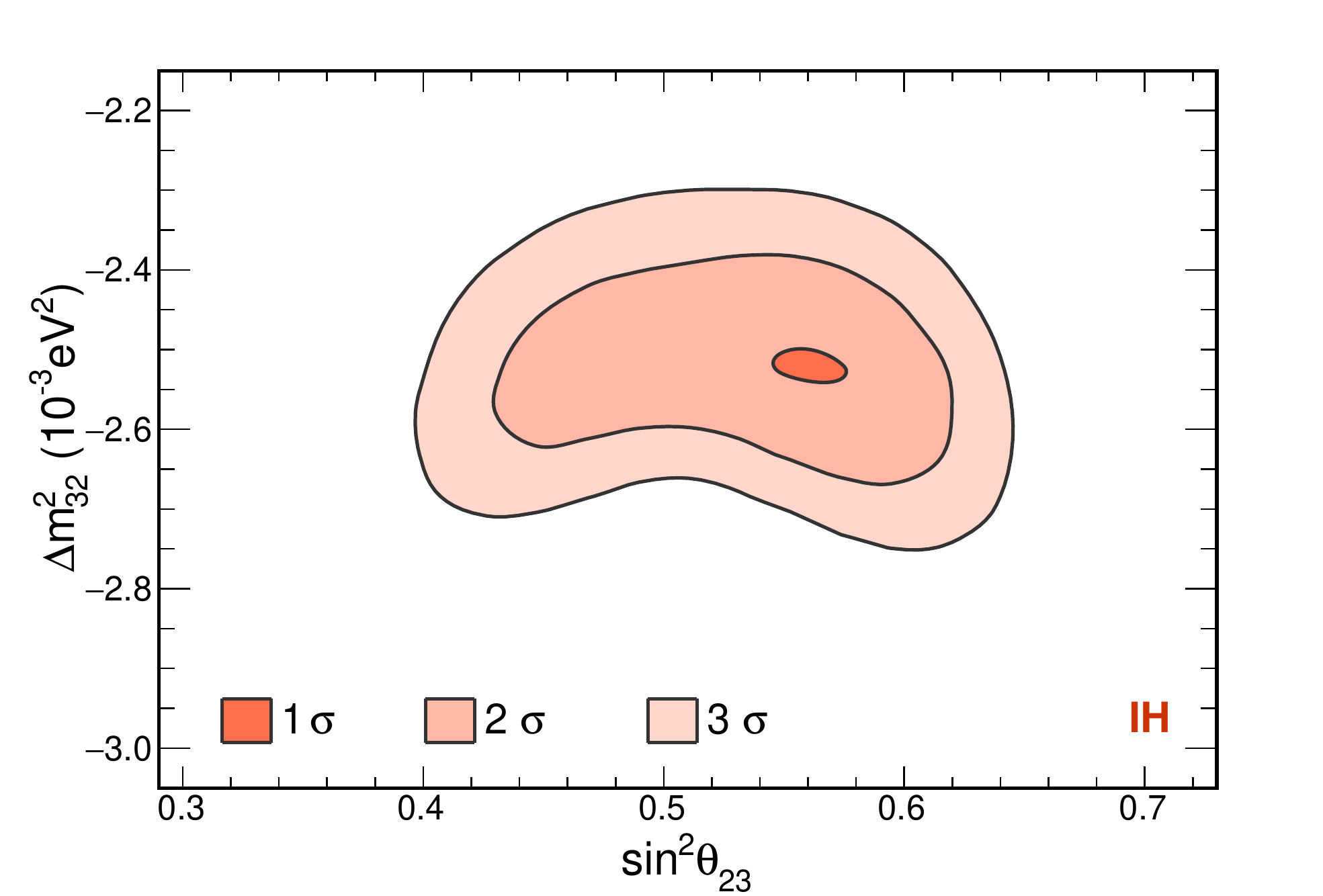}
  \caption{\label{fig:dmsqcontour}
    Regions of \dmsq{32} vs.~\sinsq{23} parameter space consistent
    with the  \nue appearance and the \numu
    disappearance data at various levels of significance. The top panel corresponds to normal
    mass hierarchy and the bottom panel to inverted hierarchy.
    The color intensity indicates the confidence level at which
    particular parameter combinations are allowed.}
\end{figure}

\begin{figure}[htb]
  \includegraphics[width=\linewidth]{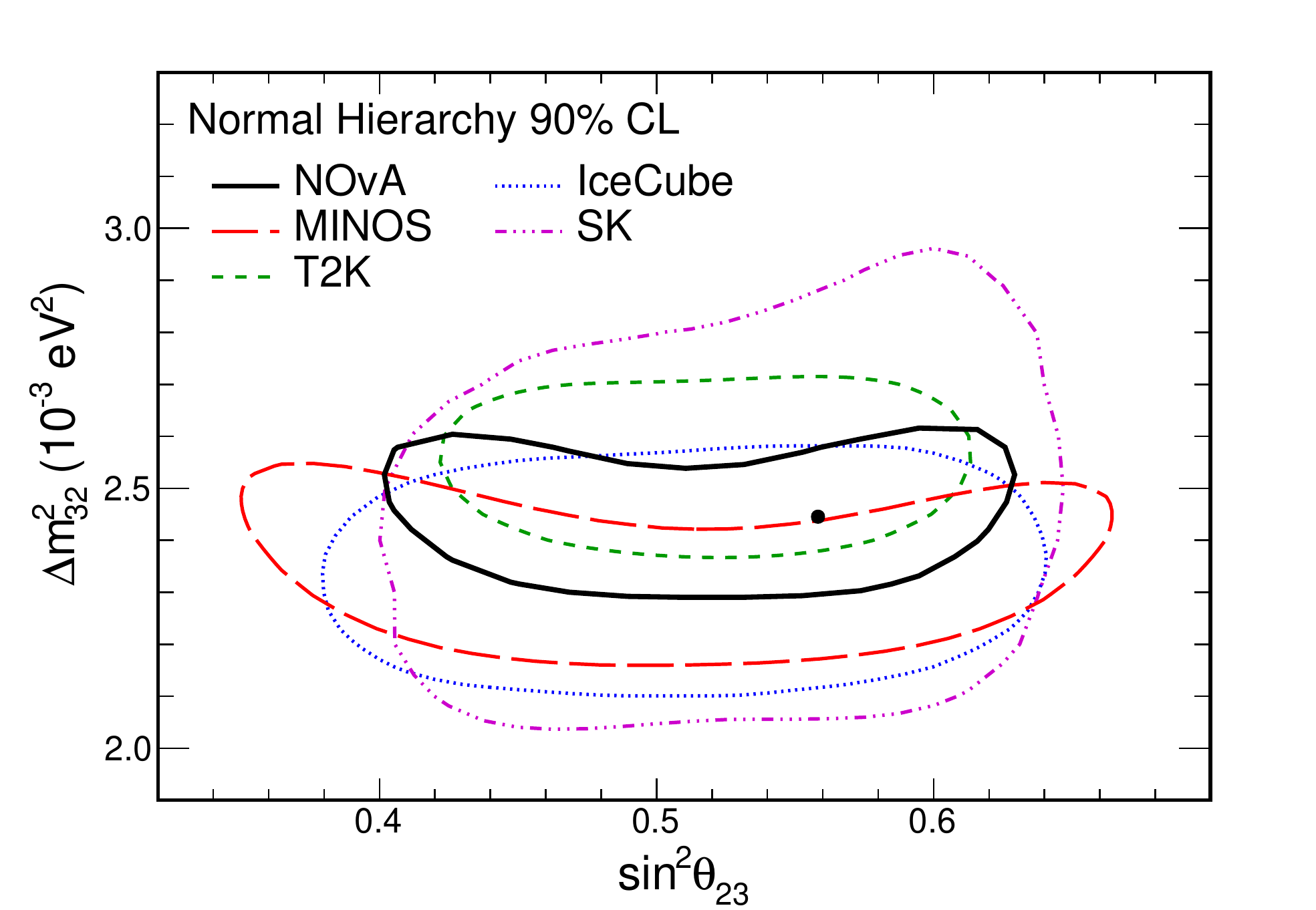}
  \caption{\label{fig:other_expt} Comparison of measured 90\% confidence level contours for \dmsq{32} vs.~\sinsq{23} for this result (black line; best-fit value, black point), T2K \cite{t2k} (green dashed), MINOS \cite{minos} (red dashed), IceCube  \cite{icecube} (blue dotted), and Super-Kamiokande \cite{superk} (purple dash-dotted).}
\end{figure}

\begin{figure}[!htb]
  \includegraphics[width=\linewidth,trim=0 45 0 0,clip]{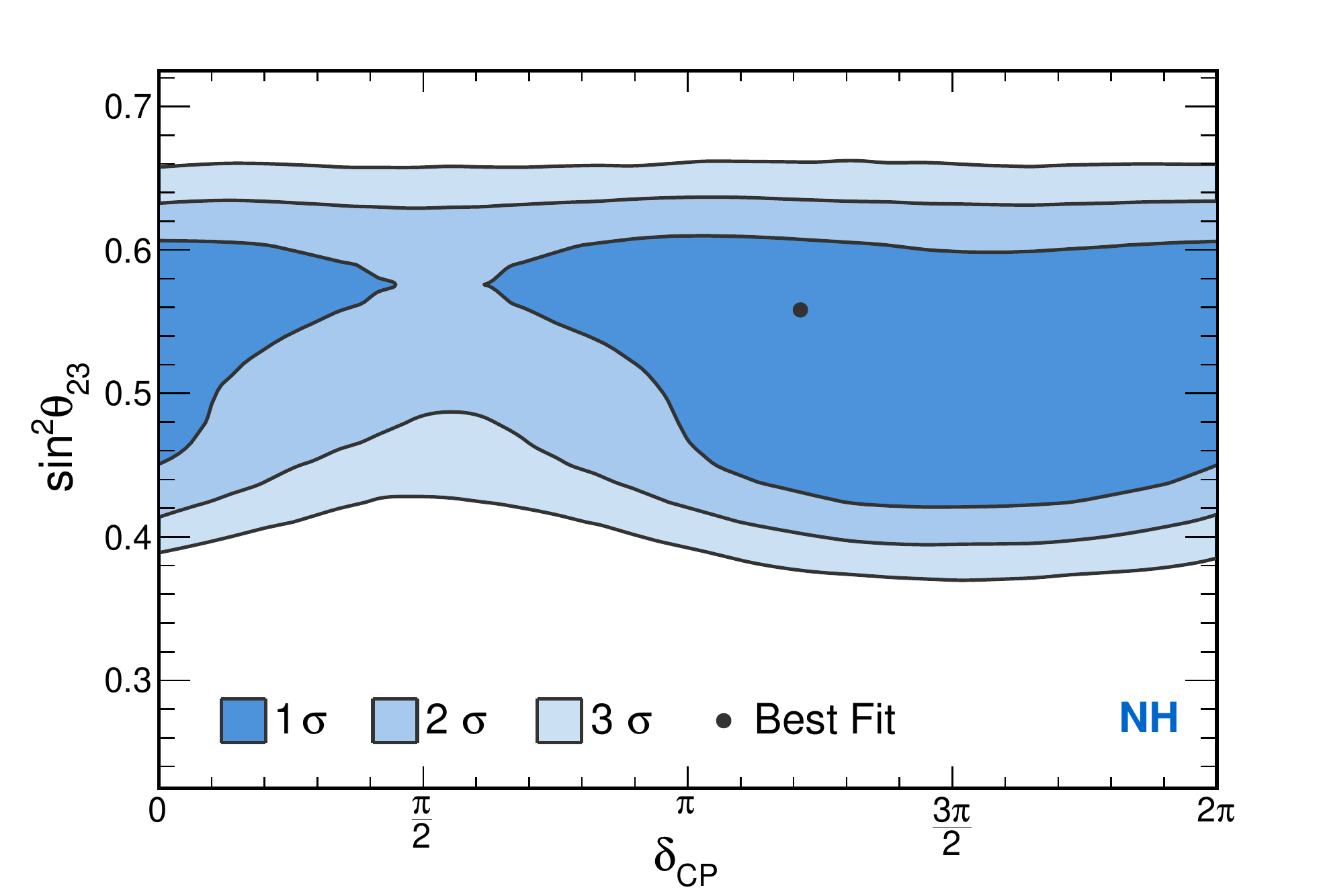}
  \vspace{-.2em}
  \includegraphics[width=\linewidth,trim=0 0 0 25,clip]{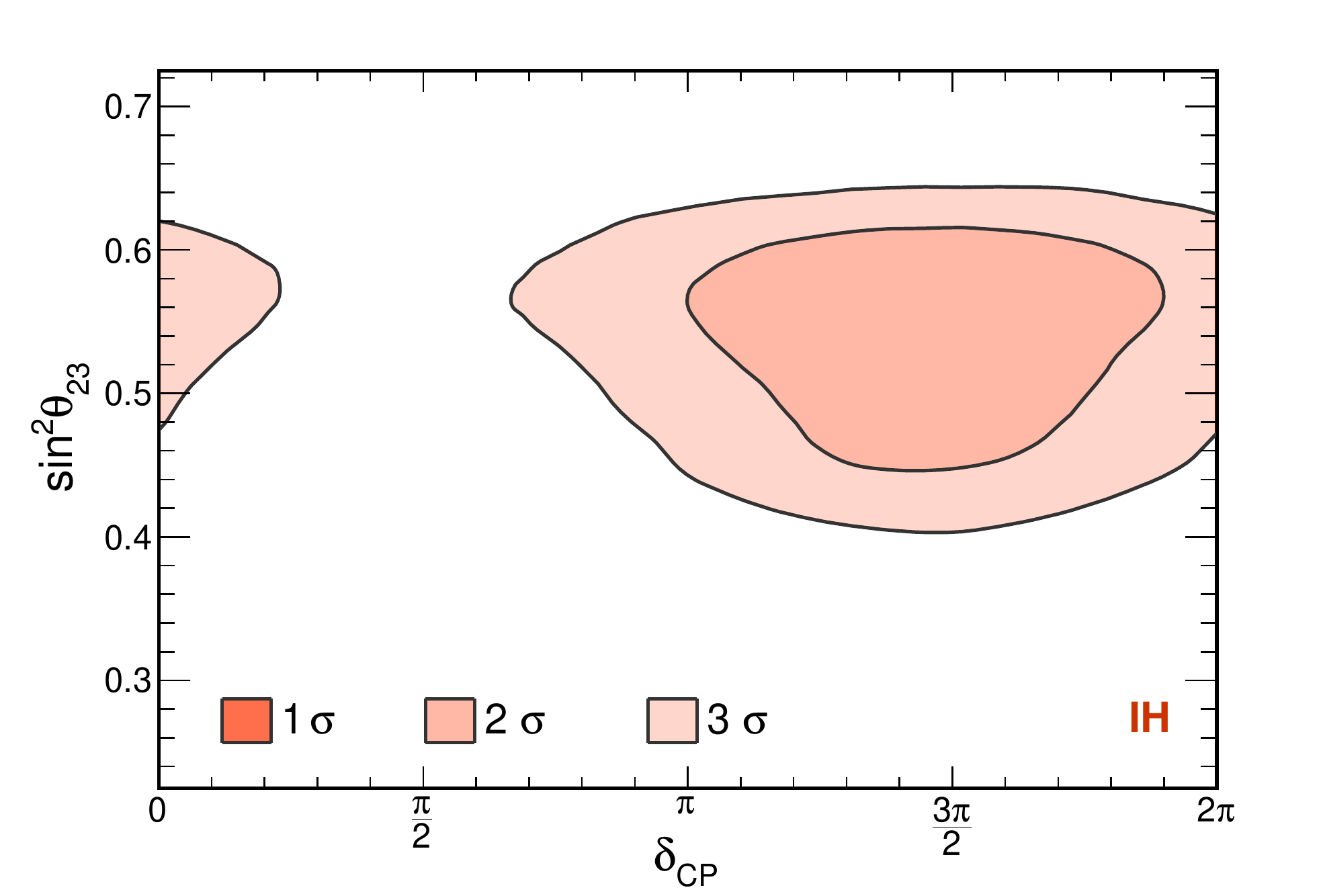}
  \caption{\label{fig:deltacontour}
    Regions of \sinsq{23} vs.~\dcp parameter space consistent
    with the  \nue appearance and the \numu
    disappearance data. The top panel corresponds to normal
    mass hierarchy ($\dmsq{32}>0$) and the bottom panel to inverted hierarchy
    ($\dmsq{32}<0$). The color intensity indicates the confidence level at which
    particular parameter combinations are allowed.}
\end{figure}

\begin{figure}[htb]
\includegraphics[width=\linewidth]{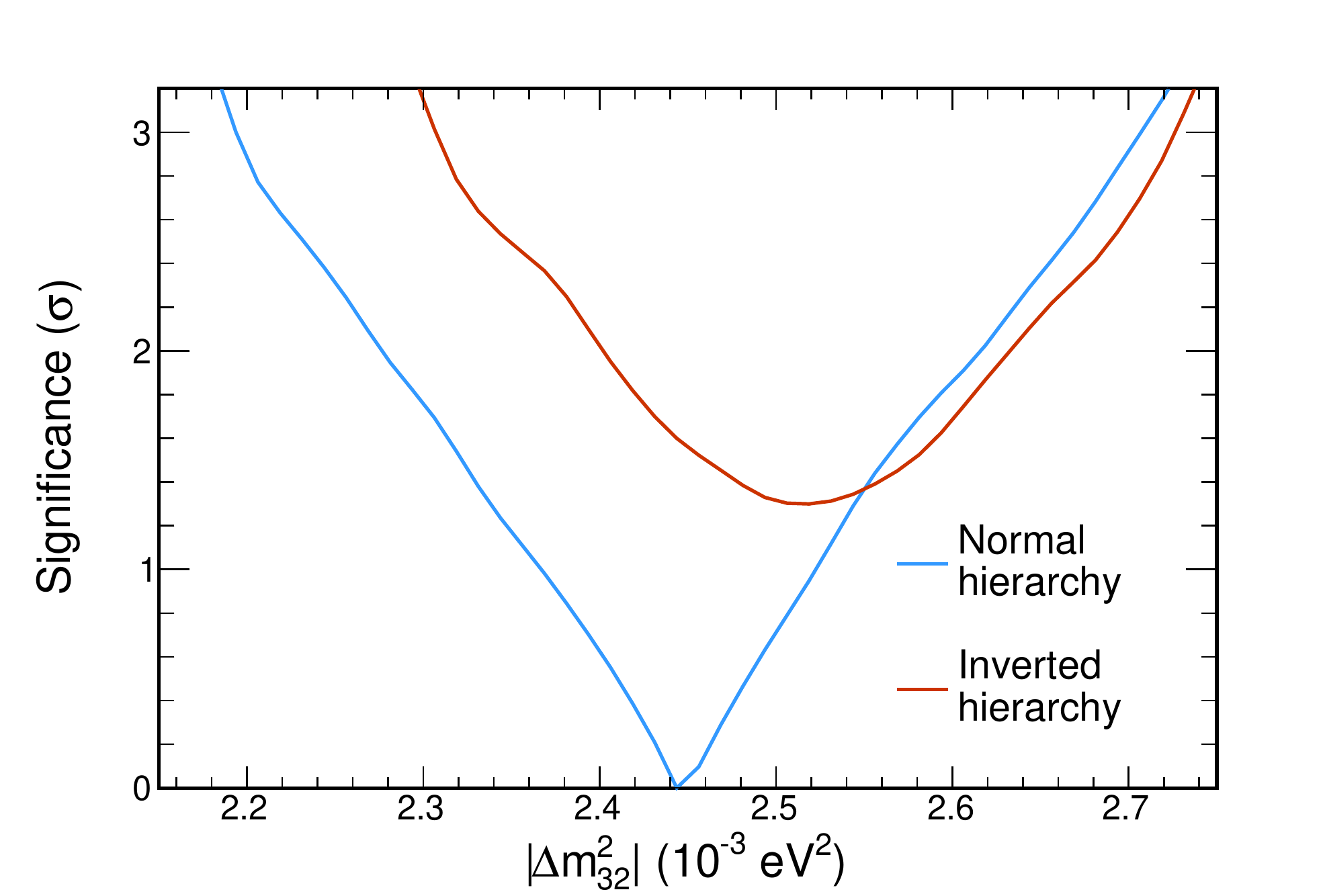}
\caption{\label{fig:dmsq}
Significance at which each value of $|\dmsq{32}|$ is disfavored
in the normal (blue, lower) or inverted (red, upper) mass hierarchy.}
\end{figure}

\subsubsection{Two dimensional contours and significance levels of single parameters}

All of the contours and significance levels that follow are constructed following the unified approach of Feldman and Cousins \cite{ref:fc}, profiling over unspecified physics parameters and systematic uncertainties. 

Figure~\ref{fig:dmsqcontour} shows the 1, 2, and 3 $\sigma$ two-dimensional contours for \dmsq{32} and \sinsq{23}, separately for each mass hierarchy.
Figure~\ref{fig:other_expt} shows a comparison of 90\% confidence level contours for these parameters in the normal mass hierarchy for NOvA, T2K \cite{t2k}, MINOS \cite{minos}, IceCube \cite{icecube}, and \mbox{Super-Kamiokande~\cite{superk}.}  
All of the experiments have results consistent with maximal mixing.  Note that the range 0.4 to 0.6 in \sinsq{23} corresponds to the range 0.96 to 1.00 in \sinsqtwo{23}, which is the variable directly measured in \numutonumu oscillations.
Figure~\ref{fig:deltacontour} shows the analogous contours to those of Fig.~\ref{fig:dmsqcontour} in \sinsq{23} and \dcp.

\begin{figure}[!tb]
\includegraphics[width=\linewidth]{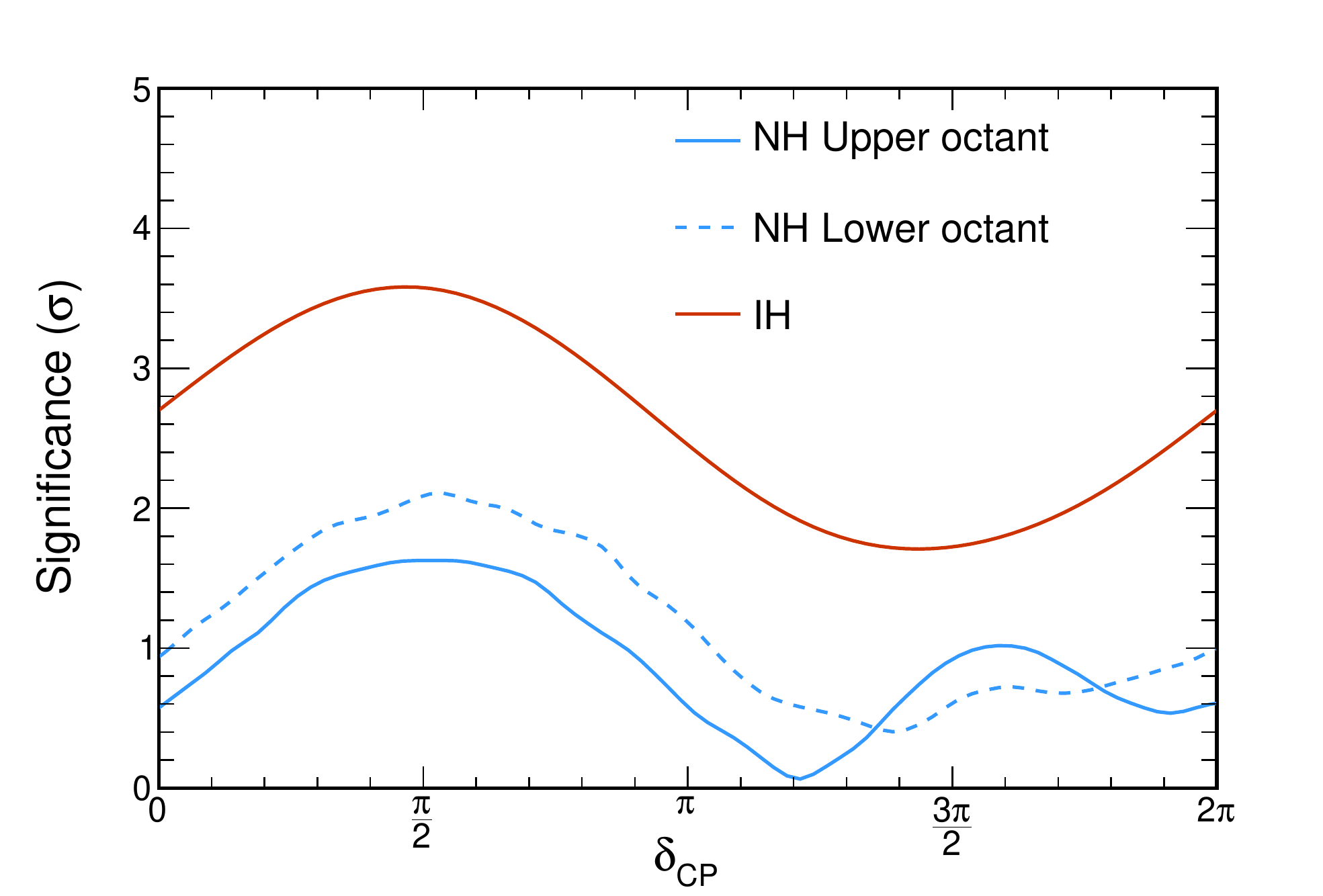}
\caption{\label{fig:delta}
Significance at which each value of \dcp is disfavored
in the normal (blue, lower) or inverted (red, upper) mass hierarchy.  The normal mass hierarchy is divided into upper (solid) and lower (dashed) $\theta_{23}$ octants corresponding to the near degeneracy in \sinsq{23}.}
\end{figure}

Figures~\ref{fig:dmsq}, \ref{fig:th23}, and \ref{fig:delta} show the significance with which values of $|\dmsq{32}|$, \sinsq{23}, and \dcp  are disfavored in the two mass hierarchies, respectively. The results in Fig.~\ref{fig:th23} differ from the ones previously reported \cite{nova_numu} in that the disfavoring of maximal mixing ($\theta_{23} = \pi/4$) has changed from 2.6 standard deviations ($\sigma$)  to \unit[0.8]{$\sigma$} in the present results. 
This change was caused by three changes, each of which moved $\theta_{23}$ closer to maximal mixing.  The largest effect was due to  new simulations and calibrations.  The two smaller effects were from  new selection and analysis procedures and from the additional  \unit[$2.80\times10^{20}$]{POT} of data included here.  The additional data taken by itself favored maximal disappearance. 
In Fig.~\ref{fig:delta} two curves are shown in the normal mass hierarchy, one for each of the $\theta_{23}$ octants, corresponding to the near degeneracy shown in Fig.~\ref{fig:th23}.  Only one curve is shown for the inverted mass hierarchy since there is only one minimum, which occurs in the upper octant. The point of minimum significance in the inverted mass hierarchy differs among the three figures because, although the $\Delta\chi^2$'s are identical (see 
Table~\ref{tab:best_fits}), the translation of $\Delta\chi^2$ to significance depends on which oscillation parameters are profiled.

Table~\ref{tab:1sigma limits} shows the \unit[1]{$\sigma$} confidence intervals for \dmsq{32}, \sinsq{23}, and \dcp in the normal mass hierarchy, corresponding to Figs.~\ref{fig:dmsq}-\ref{fig:delta}.  There are no \unit[1]{$\sigma$} confidence intervals in the inverted mass hierarchy.

\begin{table}[htb]
  \caption{\label{tab:1sigma limits} 1 $\sigma$ confidence intervals for physics parameters in the normal mass hierarchy.}
\resizebox{\linewidth}{!}{
    \begin{tabular}{p{0.35\linewidth}c}
    \hline \hline
      Parameter (units)
      & \parbox[t]{0.50\linewidth}{1 $\sigma$ interval(s)}
      
     \\ \hline \\[-14pt]
     \dmsq{32} ($10 ^{-3} \evsq$)   &     [2.37,2.52]      \\[1pt]
     \sinsq{23}    &     [0.43, 0.51] and [0.52, 0.60]      \\[1pt]
     \dcp ($\pi$) &    [0, 0.12] and [0.91, 2]       \\[1pt]

     \hline \hline
   \end{tabular}
   }
\end{table}

Finally, we have calculated the significance level for the rejection of the inverted hierarchy using the same procedure as in the above contours and confidence intervals, namely by profiling over all the other physics parameters and the systematic uncertainties.  Frequentist coverage was checked following the suggestion of Berger and Boos \cite{ref:BergerBoos}.  The entire inverted mass hierarchy region is disfavored at the 95\% confidence level.

\begin{acknowledgments}

This work was supported by the US Department of Energy; the US National Science Foundation; the Department of Science and Technology, India; the European Research Council; the MSMT CR, GA UK, Czech Republic; the RAS, RFBR, RMES, RSF and BASIS Foundation, Russia; CNPq and FAPEG, Brazil; and the State and University of Minnesota. We are grateful for the contributions of the staffs at the University of Minnesota module assembly facility and Ash River Laboratory, Argonne National Laboratory, and Fermilab. Fermilab is operated by Fermi Research Alliance, LLC under Contract No.~De-AC02-07CH11359 with the US DOE.

\end{acknowledgments}

\FloatBarrier


\begin{thebibliography}{99}
\bibitem{graphite-target} During the last 18.6\% of data collection for this analysis, a target was used which had 3 of the 48 graphite sections replaced with beryllium.  This change results in a negligible difference in the flux.
\end{thebibliography}
\end{document}